\documentclass[12pt,doublespacing]{article}
\usepackage{graphicx} 
\usepackage{amsthm, amsmath, amsfonts,mathtools, fullpage, graphicx, wrapfig, tikz, caption, subcaption}
\usepackage{natbib}
\usepackage{amssymb}
\usepackage{verbatim}
\usepackage{relsize}
\usepackage{listings}
\usepackage{xcolor}
\usepackage{colortbl}
\definecolor{darkgreen}{rgb}{0,0.69,0.31}
\definecolor{lightgreen}{rgb}{0.57,0.82,0.31}
\definecolor{yellow}{rgb}{1,1,0}
\definecolor{orange}{rgb}{1,0.75,0}
\definecolor{grey}{rgb}{0.92,0.94,0.97}
\usepackage{hyperref}
\usepackage{multirow}
\usepackage[linesnumbered,ruled,vlined]{algorithm2e}
\newcommand\independent{\protect\mathpalette{\protect\independenT}{\perp}}
\def\independenT#1#2{\mathrel{\rlap{$#1#2$}\mkern2mu{#1#2}}}

\usepackage{setspace}

\title{Generalizing Difference-in-Differences to Non-Canonical Settings: Identifying an Array of Estimands}
\author{Zach Shahn$^{1}$ and Laura A.~Hatfield$^{2}$\\
$^1$CUNY School of Public Health,  New York,  NY,  USA\\
$^2$NORC at the University of Chicago, Chicago, IL,  USA}

\begin{document}

\maketitle

\abstract{Difference-in-differences is a popular method for observational causal inference across many social science fields. It combines a two-group, two-period data structure with an assumption about counterfactual outcome evolution to identify a causal target estimand. The data structure is one group exposed to treatment in the second period only and another group that stays unexposed across both periods. The causal assumption is that the across-group difference in untreated potential outcomes would stay the same in both periods. Combining these, we can identify the average effect of treatment on the treated by taking a double difference in outcomes across group and time. In this paper, we describe an array of related combinations of data structures and causal assumptions that identify causal targets by taking a double difference. For each data structure, we imagine that we have two groups observed in (at least) two periods under some combination of treated and untreated group-periods. For instance, both groups might be treated in the second period but not the first. Then we make various assumptions about the evolution of potential outcomes. For instance, we might assume the across-group difference in potential outcomes would remain the same if both groups were treated only in the second period. Taking every possible combination of these data structures and causal assumptions, we describe the target estimand that would be identified by taking a double difference across groups and times. We consider whether it is useful or interesting and the structural justifications and plausibility checks to which we might subject the result. Using Single World Intervention Graphs (SWIGs), we describe graphical criteria for the required causal assumptions in realistic scenarios that include hierarchical data (e.g., subpopulations of individuals within treated and untreated states). We find that many more situations may be amenable to robust causal inference than previously appreciated, especially if heterogeneous treatment effects are of substantive interest. We demonstrate and apply our ideas to consider heterogeneous responses to Medicaid expansion among groups of individuals within expansion states.}

\section{Introduction}
Imagine we observe outcomes in two groups at two time periods, where some of the group-periods are treated while others are untreated.\footnote{For simplicity, we often refer to the exposure as ``treatment'', even though it might be an adverse exposure.}
Canonical Difference-in-Differences is one example: one group is treated only in the second period, and the other is untreated in both periods. 
If we assume the change in average untreated potential outcomes from the first to second period in the two groups would have been equal, then the average effect of treatment on the treated (ATT) is identified by 
a simple difference between the change (over time) in average outcomes of the treated group and the change (over time) in average outcomes of the untreated group.  

However, other arrangements of treated group-periods are possible in this two group, two period setting (see Figure \ref{fig:data_structures}). 
For example, both groups could be treated only in the second period (i.e., (b), a pre-post design), or one group treated in both periods and the other treated in neither (i.e., (c), treated-versus-control design with repeated measures).
Similarly, other parallel trends assumptions under other treatment regimes are possible (see Table \ref{tab:pts}). 
For example, we could assume the two groups' potential outcomes would evolve in parallel under a regime of `untreated in the first period and treated in the second period'. 
In fact, there is a literal array of data structures and parallel trends assumptions (Table \ref{tab:array}). 
The difference between the changes in outcomes of the two groups, which we dub the `group DiD' (gDiD) formula,
will identify different causal estimands depending on the data structure and parallel trends assumption adopted.

Are some of these combinations of estimands and identifying assumptions more useful or plausible than others?
Which assumptions are amenable to empirical checks or structural justification?
These questions motivate our inquiry into the variety of estimands and assumptions in settings with two groups observed over time.

To make things concrete, we introduce a running example, which we simplify slightly for clarity. 
\citet{kim2024synergistic} studied the relationship between country-level universal health coverage (UHC) and the outcome of childhood immunizations before and after the start of the COVID-19 pandemic. 
The two groups were countries with high and low UHC, and the two periods were pre- and post-COVID pandemic onset. 
In their gDiD analysis, \citet{kim2024synergistic} concluded, ``countries with high UHC scores prevented a 1.14\% (95\% CI: 0.39\%, 1.90\%) reduction in immunization" in the post-COVID period. 

Mapping this study to one of the panels in Figure~\ref{fig:data_structures} is a little ambiguous.
One interpretation is that high UHC is the treatment, and the high UHC group is therefore treated while the low UHC group is untreated in both periods [as in the `no pre-period' panel (c) of Figure~\ref{fig:data_structures}]. 
Another interpretation is that COVID-19 is the treatment, and both groups go from untreated in the first period to treated in the second period [as in the `pre-post' panel (b) of Figure~\ref{fig:data_structures}].
In the latter case, \citet{xu_factorial_2024} and \citet{shahn2023subgroup} showed that under a parallel trends assumption on the untreated potential outcomes across groups, gDiD identifies the \textit{difference} between the effects in the two groups. 
This is despite not identifying the effect in either group. 
In the UHC and COVID example, gDiD would therefore identify the difference between the effect of COVID on vaccination rates in high and low UHC countries.

There are numerous additional combinations of non-canonical data structures and parallel trends assumptions that lead the gDiD formula to identify a useful or interesting causal estimand (see Table \ref{tab:array}). 
Often, the required parallel trends assumption is not stronger than the canonical one and is amenable to pre-trends testing. This opens up the possibility of causal inference in scenarios such as pre-post that are conventionally considered to produce inferior causal evidence \citep{cook2007experimental}. 
The question becomes how to assess these alternative assumptions.
Numerous authors have considered the plausibility of conventional parallel trends. 
Here, we begin to bring similar reasoning to bear on the array of alternative assumptions and data structures. 

\subsection*{Related Literature}
Our paper is connected to three major strands of related research:
1) justifications for and tests of the conventional parallel trends assumption; 
2) relaxations of and alternatives to the conventional parallel trends assumption; and 
3) alternative estimands, including for heterogeneous treatment effects.

First, we ask whether justifications for the assumptions underlying canonical DiD can be adapted to other contexts.
We discuss these justifications comprehensively in Section~\ref{section_principles}, but this brief overview puts our work in context.
Parallel trends has most commonly been motivated by the parametric two-way fixed effects structural model \citep{kropko_interpretation_2020, imai_use_2021}. 
More flexible structural models \citep{ghanem_selection_2024} and causal graphs \citep{sofer2016negative,weber_assumption_2015,zhang_exploiting_2021,kim2021gain,renson2025using} have also been used to explore conditions that favor parallel trends.
A key shared insight from these studies is that parallel trends is incompatible with past outcomes directly causing treatment, which sharpens the distinction between parallel trends and related causal assumptions, like exchangeability conditional on past outcomes \citep{ding_bracketing_2019, dukes2025change}.
Many of these authors re-conceive of parallel trends as equi-confounding over time in order to apply causal graphical techniques.
\citet{ye_role_2022} similarly propose equi-confounding in two samples, each containing both exposed and unexposed units.
When one sample is a placebo that is assumed to be structurally unaffected by the treatment, the authors show that they can identify the ATT.
This resembles our expanded consideration of causal identification using parallel trends assumptions in other data structures.
Other authors have also connected parallel trends assumptions to other treatment patterns, including cases in which the treatment switches off \citep{renson2023identifying,de2023difference,shahn2022structural}.

The recent literature on empirical evidence for the parallel trends assumption has undermined the usual tests of the null hypothesis that trends are parallel in the pre-intervention period, using both theoretical and statistical arguments \citep{dette_testing_2024,freyaldenhoven_pre-event_2019,kahn-lang_promise_2020,roth_pretest_2022,bilinski_nothing_2020}.
Nonetheless, the idea that empirical evidence can support an assumption of parallel trends remains firmly entrenched in practice.
In this paper, we merely ask whether similar empirical evidence could be brought to bear on the alternative parallel trends assumptions, while assuming that all the caveats noted above still apply.

Second, among the relaxations of and alternatives to parallel trends, perhaps the simplest is to condition the assumption on baseline covariates \citep{abadie2005semiparametric,basu_constructing_2020,caetano_difference--differences_2024}. 
As we define some groups using baseline covariates, some of our parallel trends assumptions are related to conditional parallel trends.
This is the scenario described above for high versus low UHC groups before and after COVID.
The parallel trends assumption is that high and low UHC countries' trends in childhood immunization would have evolved in parallel in the absence of COVID.
Conditioning the parallel trends assumption on time-varying covariates is more complicated \citep{shahn2022structural,renson2023identifying}.
Other extensions rescue parallel trends by introducing a second comparison group that can ``net out'' the differential trends using so-called triple differences \citep{olden_triple_2022}.
However, this idea is just as frequently applied to estimate differential treatment effects in multiple groups \citep{berck_note_2016,moriya2023racial}.
In this sense, it is more closely related to the estimands we identify that are differences in treatment effects across groups. \citet{park2022universal} and \citet{tchetgen2024universal} consider replacing parallel trends with an alternative equi-confounding assumption on the odds ratio scale, which also opens up applications to settings in which the canonical assumption would fail.

Third, heterogeneous treatment effects have been the subject of many papers in the DiD literature~\citep{sun_estimating_2021,hatamyar_machine_2023,dechaisemartin_two-way_2023}.
However, they often treat heterogeneity as a nuisance, rather than a feature of the target estimand.
A special case of pre-post gDiD arises when everybody in the cohort receives treatment starting in the second time period, though different groups receive different versions of treatment. 
For example, different subgroups might receive different doses of the same treatment, as considered by \citet{callaway2024difference}. 
Then under the group `untreated' parallel trends assumption, for example, the gDiD expression identifies the difference between the effects of each treatment dose in those who received it.
The closest work to ours is probably \citet{xu_factorial_2024}, who independently noted that the combination of an untreated parallel trends assumption and a pre-post data structure can identify the difference in ATTs between two groups. 
This was also independently noted by \citet{de2023difference}.

The rest of the paper is organized as follows. In Section \ref{section_data}, we formalize the problem and describe the data structures and parallel trends assumptions of interest. In Section \ref{section_principles}, we review principles underlying the untreated parallel trends assumption of cDiD. In Section \ref{section_array}, we present our array of identification results and discuss the usefulness of the identified estimands, the plausibility of the identifying assumptions, and their possible justifications and amenability to empirical checks. In Section \ref{section_extensions}, we consider some extensions such as parallel trends conditional on covariates and bounds on conditional ATTs. In Section \ref{application}, we assess the disparate impact of ACA Medicaid expansion on uninsurance rates across racial groups under different data structures and parallel trends assumptions. In Section \ref{section_discussion}, we conclude. 

\section{Notation, Data Structures, and Assumptions}\label{section_data}
\begin{figure}[ht]
  \centering
  \begin{subfigure}[b]{0.25\textwidth}
    \centering
    \includegraphics[width=\textwidth]{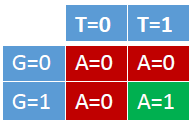}
    \caption{Canonical DiD}
  \end{subfigure}
  \hfill
  \begin{subfigure}[b]{0.25\textwidth}
    \centering
    \includegraphics[width=\textwidth]{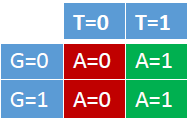}
    \caption{Pre-post}
  \end{subfigure}
  \hfill
  \begin{subfigure}[b]{0.25\textwidth}
    \centering
    \includegraphics[width=\textwidth]{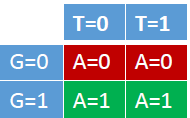}
    \caption{No pre-period}
  \end{subfigure}\\

  \begin{subfigure}[b]{0.25\textwidth}
    \centering
    \includegraphics[width=\textwidth]{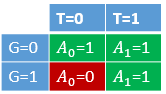}
    \caption{Treatment turns on vs. treated}
  \end{subfigure}
   \hfill
 \begin{subfigure}[b]{0.25\textwidth}
    \centering
    \includegraphics[width=\textwidth]{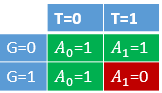}
    \caption{Treatment turns off vs. treated}
  \end{subfigure}
    \hfill
     \begin{subfigure}[b]{0.25\textwidth}
    \centering
    \includegraphics[width=\textwidth]{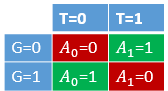}
 \caption{Treatment turns off vs. untreated}
  \end{subfigure}\\
 \begin{subfigure}[b]{0.25\textwidth}
    \centering
    \includegraphics[width=\textwidth]{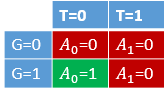}
 \caption{Crossover}
  \end{subfigure}
    \hfill
   \begin{subfigure}[b]{0.25\textwidth}
    \centering
    \includegraphics[width=\textwidth]{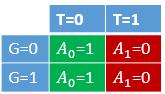}
 \caption{Pre-post switch off}
  \end{subfigure}
  \caption{Data structures for which gDiD identifies useful causal quantities given a group parallel trends assumption (See Table \ref{tab:pts}) and Consistency (\ref{eq:consistency}).
  Panel (a) canonical DiD setting in which all units are untreated in the first period, and group $G$ indexes the second period treatment, i.e., $A_0=0$, $A_1=G$. 
  Panel (b) pre-post setting in which no units are treated in the first period, all are treated in the second period, and $G$ is defined by baseline covariates. 
  Panel (c) no pre-period setting in which group $G$ indexes the treatment assignment received at both time periods, i.e., $A_0=A_1=G$. 
  Panel (d) treatment switches on in one group and is always on in the other. 
  Panel (e) treatment switches off in one group and remains on in the other. 
  Panel (f) treatment switches off in one group and is always off in the other. 
  Panel (g) `crossover' setting in which one group switches treatment from on to off and the other from off to on. 
  Panel (h) pre-post setting in which treatment switches off for all units, and $G$ is defined by baseline covariates.}
  \label{fig:data_structures}
\end{figure}

\clearpage
To formalize the problem, assume the data comprise independent and identically distributed realizations of the random variable $O=(G,A_0,Y_{0},A_1,Y_{1})$, where $G$ denotes a binary group indicator, $A_t$ denotes treatment at time $t$, and $Y_t$ denotes outcome at time $t$ for $t\in\{0,1\}$. 
Figure \ref{fig:data_structures} uses this notation to illustrate possible treatment assignments in each group-period.
For instance, Panel (a) shows the cDiD data structure where $A_0=0$ and $A_1=G$, Panel (b) shows the pre-post design where $A_0=0$ and $A_1=1$ for both groups, and Panel (c) shows the `group as treatment' setting where $A_t=G$ for both periods.  

The coding of $A_t=1$ and $A_t=0$ is somewhat arbitrary but the conventional parallel trends assumption is made with respect to untreated outcomes, so it does matter. 
When we say that a group-period has treatment assignment $A=0$, we may mean any of the following:
\begin{enumerate}
\item[a)] Unaffected by $A=1$ (as in \citet{xu_factorial_2024}),
\item[b)] Never exposed to $A=1$, or
\item[c)] Responses to previous exposure to $A=1$ have worn off.
\end{enumerate}
Consider a cDiD study estimating the effect of repealing a law. In such a study, the `no law' condition might be deemed `treated'. 

Now let $Y_0(a_0)$ and $Y_1(a_0,a_1)$ denote the potential outcomes at times $t=0$ and $t=1$, respectively, under the regime $(a_0,a_1)$.
We make the following consistency assumption:
\begin{equation}\label{eq:consistency}
\textbf{Consistency: } (Y_0(a_0),Y_1(a_0,a_1)) = (Y_0,Y_1) \text{ when }(A_0,A_1)=(a_0,a_1). 
\end{equation}
Implicit here is a `no anticipation' assumption that the future does not affect the past and `no interference' assumption that one unit's treatment assignment does not impact another unit's outcome.
Later, we discuss adding an additional `no carryover effects' assumption, but for now, note that the $t=1$ observed outcome is the potential outcome given the regime in \emph{both} periods. 

We consider the contrast of average outcomes across both group and time, which we call the gDiD,\begin{equation}\label{eq:gdid}
E[Y_1-Y_0|G=1] - E[Y_1-Y_0|G=0]\;,
\end{equation}
and explore the data structures and causal assumptions under which this identifies a meaningful causal estimand.
We limit our attention to causal assumptions that resemble parallel trends. 
For example, the parallel trends assumption typically invoked to justify cDiD is that the two groups would experience equal outcome changes under the regime $(a_0,a_1) = (0,0)$, that is,
\begin{equation}\label{gpt}
    E[Y_1(0,0)-Y_0(0)|G=1] = E[Y_1(0,0)-Y_0(0)|G=0].
\end{equation}

What parallel trends assumptions might we make under \emph{other} treatment regimes that would be useful for different data structures?
With 4 possible regimes, there are 16 possible combinations of regimes in two groups (i.e., $4 \times 4$), but many of these are quite implausible.
Therefore, we generally only consider assumptions where the regimes in both groups match, that is, of the form
$$
E[Y_1(a_0,a_1)-Y_0(a_0)|G=1] = E[Y_1(a_0,a_1)-Y_0(a_0)|G=0],
$$ 
with one exception (discussed below). Note that we limit our attention to cases where the regime in the second period potential outcome includes the same first period treatment as the first period potential outcome. 
That is, we do not consider trends of the form $E[Y_1(a_0, a_1) - Y_0(a_0')|G = g]$ where $a_0\neq a_0'$.
Table~\ref{tab:pts} enumerates the five parallel trends assumptions we do consider. 

The exception to the general `matching regimes' rule above is the `never switch' parallel trends assumption. 
This assumption states that trends are parallel under a regime in which neither group changes treatment in the $t=1$ period from its observed treatment in the $t=0$ period. The implications of this assumption can depend on the observed data structure. Note that under the cDiD data structure, for example, the `untreated' and `never switch' parallel trends assumptions actually coincide. In that data structure, because nobody is treated at the first time point, not switching corresponds to not treating at both time points. However, under the data structure from panel (f) of Figure \ref{fig:data_structures}, these two assumptions differ. The `never switch' assumption states that the $G=1$ trend would have been equal to the observed $G=0$ trend had $G=1$ continued to receive treatment at the $t=1$ time point. The `untreated' assumption states that the $G=1$ trend would have been equal to the observed $G=0$ trend had $G=1$ never received treatment. The `never switch' parallel trends assumption is an example of a parallel trends assumption under a \textit{dynamic} treatment regime, i.e., a treatment strategy where the treatment at the second time point depends on history, specifically the treatment at the first time point.

\begin{table}
    \centering 
    \begin{tabular}{cc}
        Untreated & $E[Y_1(0,0)-Y_0(0)|G=1] = E[Y_1(0,0)-Y_0(0)|G=0]$\\
         Treated & $E[Y_1(1,1)-Y_0(1)|G=1] = E[Y_1(1,1)-Y_0(1)|G=0]$\\
         Switch on& $E[Y_1(0,1)-Y_0(0)|G=1] = E[Y_1(0,1)-Y_0(0)|G=0]$\\
         Switch off& $E[Y_1(1,0)-Y_0(1)|G=1] = E[Y_1(1,0)-Y_0(1)|G=0]$\\
         Never Switch&$E[Y_1(A_0,A_0)-Y_0(A_0)|G=1] = E[Y_1(A_0,A_0)-Y_0(A_0)|G=0]$ \\
    \end{tabular}
    \caption{Group parallel trends assumptions under a variety of possible treatment regimes}
    \label{tab:pts}
\end{table}

\section{Review of principles underlying canonical DiD}\label{section_principles}
Due to the popularity of cDiD, much prior work has considered structural and empirical justifications for the `untreated' parallel trends assumption (row 1 of Table \ref{tab:pts}) in the cDiD setting (panel (a) of Figure \ref{fig:data_structures}).
We will later consider whether similar justifications would apply to other parallel trends assumptions for other data structures. 
To that end, in this section we offer a simplified discussion of some principles that are often invoked to justify cDiD. 
We will refer back to these when we discuss our gDiD alternatives in the next section.   

\subsection{Assumptions about treated vs untreated potential outcomes}\label{section_treatment_fundamentals}
The first two principles are largely implicit in discussions of causal identifying assumptions for cDiD.
They set the terms of debate: make a parallel trends assumption that concerns only untreated potential outcomes and apply it to a setting in which treatment may switch on but not off.
Because we will consider parallel trends assumptions that also concern treated potential outcomes and more diverse data structures, we find it instructive to make these principles explicit.

Untreated potential outcomes are privileged as a ``natural" state, or one in which equilibrium has been achieved.
A novel treatment or exposure, by contrast, may alter the equilibrium
and `contaminate' later group-period observations that are nominally `untreated'. 
Thus, we ideally apply cDiD when the treatment has not occurred in any units at any time, even prior to $t=0$, before $t=1$. 
Assessing this requires deep knowledge of the policy landscape-- we would need a comprehensive scan of the policy history of our potential comparison units for exposures in the past that might still be affecting outcomes.
In fact, many DiD studies fail to thoroughly search for policies related to the focal policy that might have recently turned on in the pool of potential controls.


The following two additional principles propose underlying causal structural models that would justify untreated parallel trends under the cDiD data structure.

\subsection{Parametric structural models}
A nearly ubiquitous motivating model for cDiD in the econometrics literature says that untreated potential outcomes are generated by
\begin{equation}\label{eq:fixed_effects}
    Y_{it}(0,0) = \lambda_t + f(U_i) + \epsilon_{it},
\end{equation}
where $U_i$ are unit-specific time-invariant characteristics (aka ``unit fixed effects''), $\lambda_t$ are time-varying intercepts (aka ``time fixed effects''), and $\epsilon_{it}$ are idiosyncratic errors such that $E[\epsilon_{it}|A_1,U_i]=0$.

Eq.~(\ref{eq:fixed_effects}) is the so-called ``two-way fixed effects'' model, where the ``two-way'' refers to the unit and time fixed effects.
It is trivial to show that if this structural model is true, `untreated' parallel trends holds.
This model also encodes several important features.
For instance, $f(U_i)$ does not depend on $t$, so outcomes that have common causes with treatment $A_1$ are impacted similarly by these common causes at each time (on an additive scale). 
While the distribution of $U_i$ may differ in the treated and untreated groups, the time-invariant $f(U_i)$ cancels out when we take differences within groups in the gDiD formula. 
Moreover, the impact of time-dependent `shocks' $\lambda_t$ are the same for all units. 
There are no restrictions on the associations between the errors $\epsilon_{it}$ at different times, so untreated outcomes may have arbitrary associations across time. 

Next, we consider a different kind of causal structural model that can justify the `untreated' parallel trends assumption.

\subsection{Graphical structural methods}
We begin by expressing the untreated parallel trends assumption as,
\begin{equation}\label{eq:cdid_indep}
E\left[\Delta_1(0,0)|A=1\right] = E\left[\Delta_1(0,0)|A=0\right]
\end{equation}
where $\Delta_1(0,0) = Y_1(0,0)-Y_0(0)$.
If $A$ and $\Delta_1(0,0)$ are independent, this mean equality will also hold.
Independence is stronger than required, but we argue it would be unusual for the mean equality to hold without independence.
More important, independence relationships can be depicted using causal graphs, whereas mean equality on an additive scale is extra-graphical.



We illustrate the graphical analysis of this parallel trends assumption using Single World Intervention Graphs
(SWIGs) \citep{richardson2013single}.
These are directed acyclic graphs that include both interventions and counterfactual outcomes,
which enables them to encode conditional independence relationships between observed and counterfactual variables. 
SWIGs allow us to reason about ``what would happen'' under different treatment regimes.
Split nodes depict variables on which we intervene to set the values for a particular regime.
The left side represents a variable's observed value and the right side its value under our intervention. 
Conditioning is represented by a rectangular box for a random variable's node.

\begin{figure}[ht]
    \centering
    \includegraphics[width=0.49\linewidth]{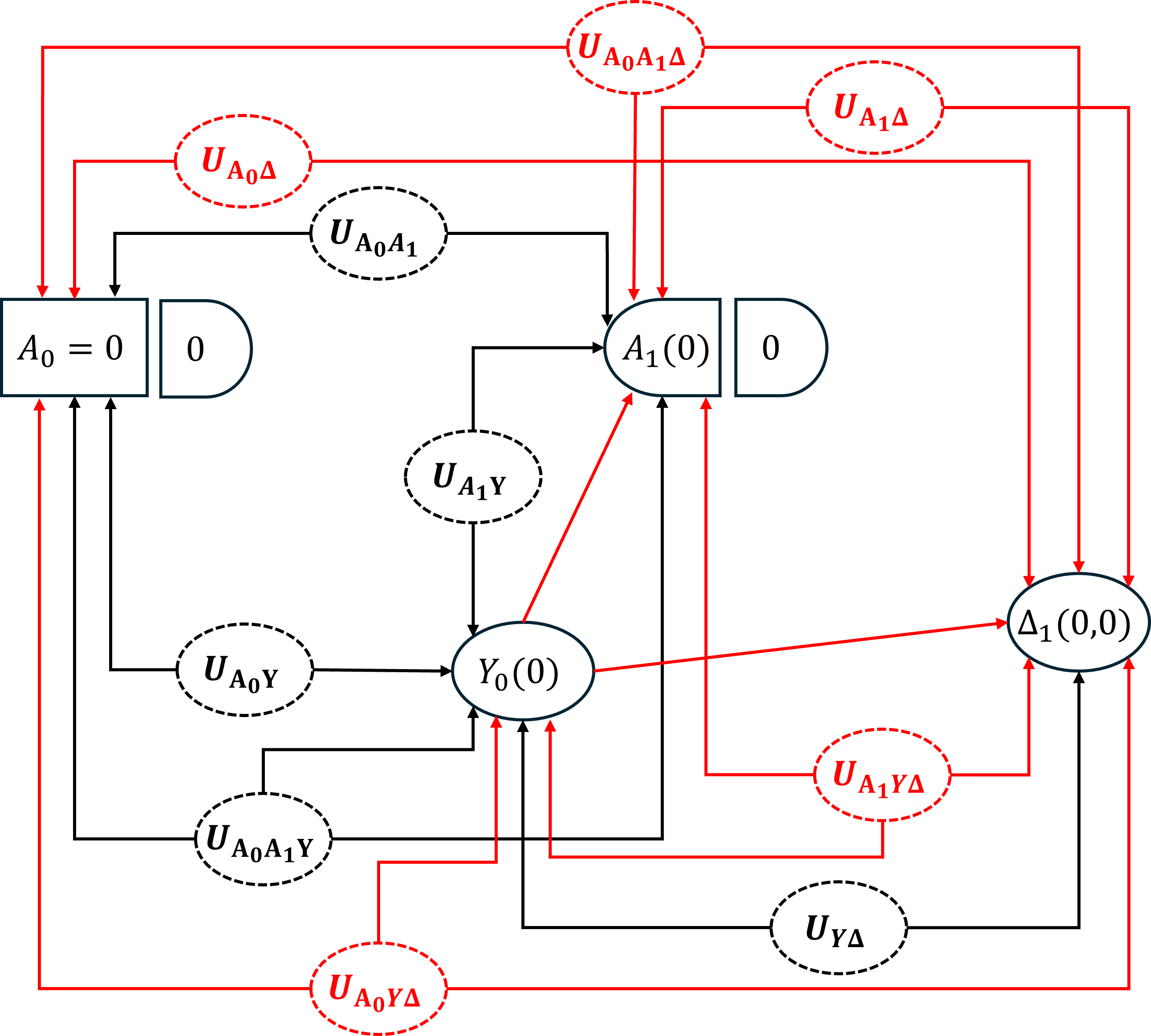} \hfill
    \includegraphics[width=0.49\linewidth]{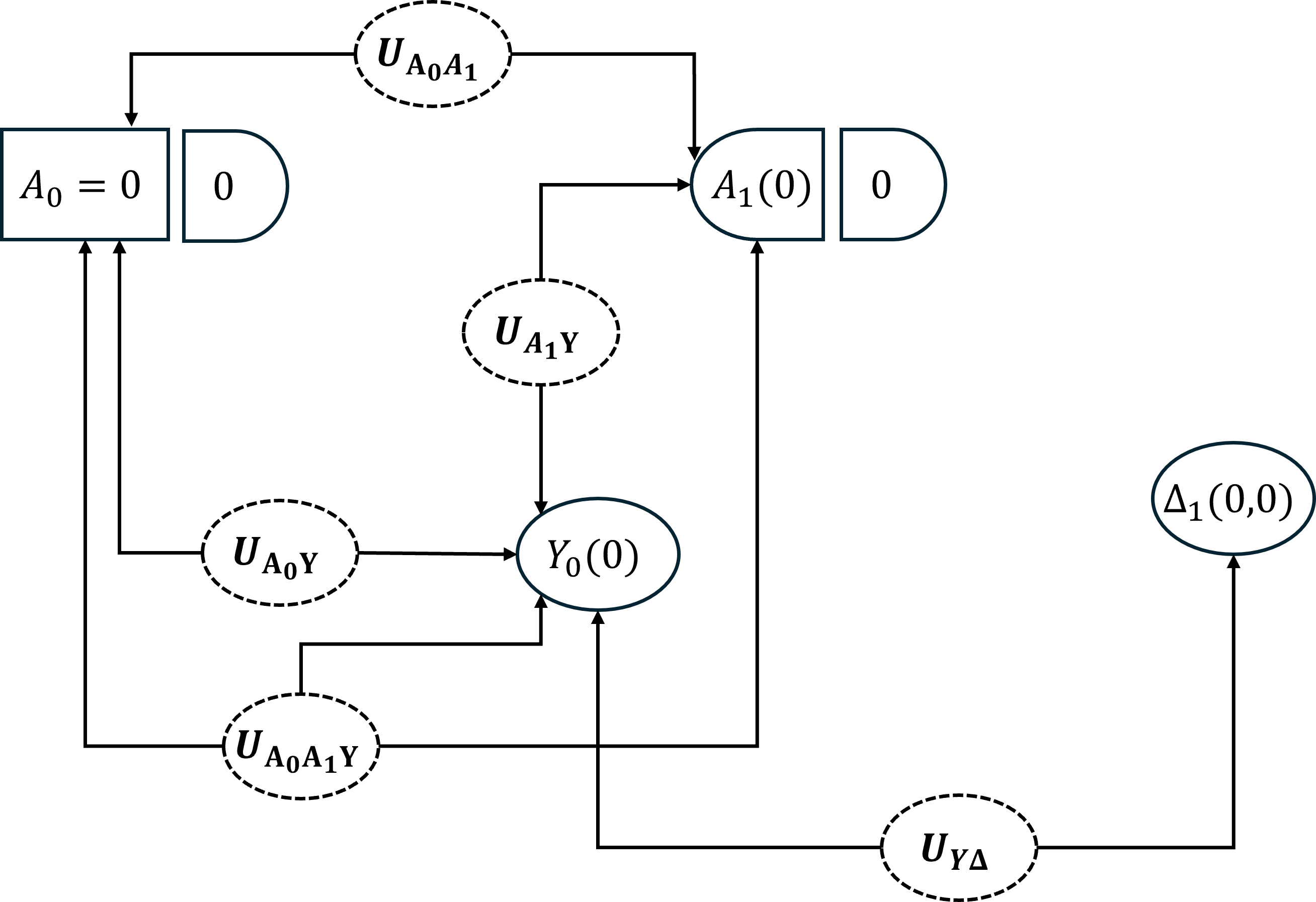}
    \caption{A Single World Intervention Graph illustrating the canonical DiD design before (left) and after (right) making the independence version of the parallel trends assumption.}
    \label{fig:cdid_swig}
\end{figure}

The right panel of Figure \ref{fig:cdid_swig} is a SWIG that depicts the parallel trends assumption given in Eq.~(\ref{eq:cdid_indep}).
How have we arrived at this graph?
We began with the ``maximal graph'' in the left panel of Figure \ref{fig:cdid_swig} then pruned it to get the graph representing the independence we need.
It encodes several important assumptions about this setting.
First, two assumptions are about unobservable common causes that we assume \emph{do} exist.
We assume there \emph{are} unobserved common causes of treatment assignment and baseline outcomes, 
otherwise we would not need DiD at all.
We also assume there are unobserved common causes of baseline outcomes and untreated outcome trends.
Otherwise, the time series of outcomes would be a random walk, in which case it can be shown that parallel trends implies partial conditional exchangeability (i.e. $A\independent Y_1(0)|Y_0$) \citep{dukes2025change}, again making DiD unnecessary. 
These two assumptions constrain the ways we can possibly make the graph free of backdoor paths between $A_1$ and $\Delta_1$.
We must assume there are no unobserved common causes of treatment assignment and untreated outcome trends,
which removes all the unobserved nodes marked in red (those with arrows into both treatment and the untreated trend) in the left panel of Figure \ref{fig:cdid_swig}.
We must also assume that baseline outcomes do not cause either treatment assignment or untreated outcome trends,
which removes the two arrows out of $Y_0$, also shown in red in the left panel.
Section~\ref{section_swigs} of the appendix details our process of constructing this SWIG.

Above, we have used SWIGs to derive certain general \textit{implications} of parallel trends, in the form of necessarily missing arrows if parallel trends holds. One might also wish to use a SWIG to determine whether parallel trends holds in a particular application. However,
using SWIGs in this way is not as straightforward as it is for assessing exchangeability. 
The absence of arrows from $U_{A_1Y}$ to $\Delta(0,0)$, for example, is scale-dependent.
Supposing that $U_{A_1Y}$ directly causes both $Y_0$ and $Y_1$ (as would be typical), it can only fail to cause $\Delta(0,0)$ (on average) if its effects on $Y_0$ and $Y_1$ cancel out in their difference \textit{at the chosen scale}. 
To assess whether such canceling out actually occurs would require detailed structural knowledge of the data generating process.  \citet{ghanem_selection_2024} consider what sorts of assumptions on the structural equations might justify parallel trends.



\subsection{Tests of parallel trends}
The last principle is a sort of meta-principle about the causal assumption, its amenability to empirical ``tests'', rather than a structural justification.
A common auxiliary principle in cDiD studies is that the `untreated' parallel trends assumption in row 1 of Table \ref{tab:pts} holds in some set of historical pre-intervention periods, $t \in\{-K,\ldots,-1\}$, 
\begin{equation}\label{eq:pre-trends}
    E[Y_t(\bar{0}_t)-Y_{t-1}(\bar{0}_{t-1})|A_1=1]=E[Y_t(\bar{0}_t)-Y_{t-1}(\bar{0}_{t-1})|A_1=0]\;.
\end{equation}
In this expression, $Y_t(\bar{0}_t)$ is the potential outcome under a regime that contains only $0$ values through time $t$. Notice that if we extend the time index to these new periods, the two-way fixed effects structural model above implies that this ``extended parallel trends'' assumption holds.
Even without a compatible structural model, we can simply invoke this new assumption to justify looking for empirical evidence of parallel trends in the historical period.
The idea is that this will muster support for the required causal assumption, which must hold from the pre- to post-intervention period and is therefore untestable.
Thus, another principle on which we will judge our combinations of parallel trends assumptions and data structures is this: can we look for empirical evidence of the required parallel trends assumption, perhaps by extending it to additional periods?
Of course, all the caveats noted above about the usefulness of testing for parallel pre-trends apply \citep{bilinski_nothing_2020, freyaldenhoven_pre-event_2019, kahn-lang_promise_2020, roth_pretest_2022}.

With this grounding in the familiar case of canonical DiD, which combines the data structure from Panel (a) in Figure~\ref{fig:data_structures} with the `untreated' parallel trends assumption from row 1 of Table~\ref{tab:pts}, we next introduce the full array of combinations of data structures and parallel trends assumptions.

\section{The Array of Identification Results}\label{section_array}

\begin{table}[ht]
    \centering
    \tiny
    \begin{tabular}{>{\raggedright\arraybackslash}p{0.1\textwidth}|
    >{\raggedright\arraybackslash}p{0.15\textwidth}|
    >{\raggedright\arraybackslash}p{0.15\textwidth}|
    >{\raggedright\arraybackslash}p{0.15\textwidth}|
    >{\raggedright\arraybackslash}p{0.15\textwidth}|
    >{\raggedright\arraybackslash}p{0.15\textwidth}|}
        \hline
        & \textbf{Untreated} & \textbf{Treated} & \textbf{Switch on} & \textbf{Switch off} & \textbf{Never switch} \\
        \hline
        \textbf{a) Canonical DiD} & 
        \cellcolor{darkgreen}
        $E[Y_1 (0,1) - Y_1 (0,0) \mid G = 1]$ & 
        \cellcolor{grey}
        $(E[Y_1 - Y_1 (1,1) \mid G=1] - E[Y_0 - Y_0 (1) \mid G=1]) - 
         (E[Y_1 - Y_1 (1,1) \mid G=0] - E[Y_0 - Y_0 (1) \mid G=0])$ & 
         \cellcolor{darkgreen}
        $E[Y_1 (0,0) - Y_1 (0,1) \mid G = 0]$ & 
        \cellcolor{grey}
        $(E[Y_1 (0,1) - Y_1 (1,0) \mid G=1] - E[Y_0 (0) - Y_0 (1) \mid G=1]) - 
         (E[Y_1 (0,0) - Y_1 (1,0) \mid G=0] - E[Y_0 (0) - Y_0 (1) \mid G=0])$ &
         \cellcolor{darkgreen}
        $E[Y_1 (0,1) - Y_1 (0,0) \mid G = 1]$ \\
        \hline
        \textbf{b) Pre-post} & 
        \cellcolor{lightgreen}
        $E[Y_1(0,1) - Y_1(0,0)\mid G=1] - E[Y_1(0,1)-Y_1(0,0) \mid G=0]$ &
        \cellcolor{yellow}
        $E[Y_0 (0) - Y_0(1)\mid G=1] - E[Y_0(0)-Y_0(1)\mid G=0]$ &
        \cellcolor{grey}
        $0$	&
        \cellcolor{yellow}
        $E[Y_1(1) - Y_1(0)\mid G=1]+E[Y_0(1)-Y_0(0)\mid G=1) -
        (E[Y_1(1) - Y_1(0)\mid G=0]+E[Y_0(1)-Y_0(0)\mid G=0])$ &
        \cellcolor{lightgreen}
        $E[Y_1(0,1) - Y_1(0,0)\mid G=1]-E[Y_1(0,1)- Y_1(0,0)\mid G=0]$ \\
        \hline
        \textbf{c) No pre-period} & 
        \cellcolor{lightgreen}
        $E[Y_1 (1,1)-Y_1 (0,0)\mid G=1]-E[Y_0 (1)-Y_0 (0)\mid G=1] $ & 
        \cellcolor{lightgreen}
        $E[Y_1 (0,0)-Y_1 (1,1)\mid G=0]-E[Y_0 (0)-Y_0 (1)\mid G=0] $ &
        \cellcolor{yellow}
        $E[Y_1 (0,1)-Y_1 (0,0)\mid G=0]-E[Y_0 (1)-Y_0 (0)\mid G=1] $ & 
        \cellcolor{yellow}
        $E[Y_1 (1)-Y_1 (0)\mid G=1]- E[Y_0 (1)-Y_0 (0) \mid G=0] $ &
        \cellcolor{grey}
        $0$ \\
        \hline
        \textbf{d) Treatment turns on vs treated} &
        \cellcolor{grey}
        $E[Y_1 (0,1)-Y_1 (0,0)\mid G=1]-(E[Y_1 (1,1)-Y_1 (0,0)\mid G=0]-E[Y_0 (1)-Y_0 (0)\mid G=0]) $ & 
        \cellcolor{orange}
        $E[Y_0 (1)-Y_0 \mid G=1] $ & 
        \cellcolor{darkgreen}
        $E[Y_1 (1,1)-Y_1 (0,1) \mid G=0]$ &
        \cellcolor{grey}
        $(E[Y_1 (0,1)-Y_1 (1,0)\mid G=1]-E[Y_0 (0)-Y_0 (1)\mid G=1])- E[Y_1 (1,1)-Y_1 (1,0)\mid G=0] $ & 
        \cellcolor{darkgreen}
        $E[Y_1 (0,1)-Y_1 (0,0) \mid G=1] $ \\        			
        \hline
        \textbf{e) Treatment turns off vs treated} & 
        \cellcolor{grey}
        $(E[Y_1 (1,0)-Y_1 (0,0)\mid G=1]-E[Y_0 (1)-Y_0 (0)\mid G=1])-(E[Y_1 (1,1)-Y_1 (0,0)\mid G=0]-E[Y_0 (1)-Y_0 (0)\mid G=0]) $ & 
        \cellcolor{darkgreen}
        $E[Y_1 (1,0)-Y_1 (1,1)\mid G=1] $ & 
        \cellcolor{grey}
        $(E[Y_1 (1,0)-Y_1 (0,1)\mid G=1]-E[Y_0 (1)-Y_0 (0)\mid G=1])-E[Y_1 (0,0)-Y_1 (0,1)\mid G=0] $ & 
        \cellcolor{darkgreen}
        $E[Y_1 (1,0)-Y_1 (1,1)\mid G=0] $ & 
        \cellcolor{darkgreen}
        $E[Y_1 (1,0)-Y_1 (1,1)\mid G=1] $ \\	
        \hline
        \textbf{f) Treatment turns off vs untreated} & 
        \cellcolor{lightgreen}
        $E[Y_1 (1,0)-Y_1 (0,0)\mid G=1]-E[Y_0 (1)-Y_0 (0)\mid G=1] $ & 
        \cellcolor{grey}
        $E[Y_1-Y_1 (1,1)\mid G=1]-(E[Y_1-Y_1 (1,1)\mid G=0]-E[Y_0-Y_0 (1)\mid G=0]) $ & 
        \cellcolor{grey}
        $(E[Y_1 (1,0)-Y_1 (0,1)\mid G=1]-E[Y_1 (1)-Y_0 (0)\mid G=1])-E[Y_1 (0,0)-Y_1 (0,1)\mid G=0] $ & 
        \cellcolor{orange}
        $E[Y_0 (0)-Y_0 (1)\mid G=0] $ & 
        \cellcolor{darkgreen}
        $E[Y_1 (1,0)-Y_0 (1,1)\mid G=1]$\\
        \hline
        \textbf{g) Crossover} & 
        \cellcolor{yellow}
        $E[Y_1 (0,1)-Y_1 (0,0)\mid G=0]-E[Y_0 (1)-Y_0 (0)\mid G=1] $ &
        \cellcolor{yellow}
        $E[Y_0 (0)-Y_0 (1)\mid G=0]-E[Y_1 (1,0)-Y_1 (1,1)\mid G=1] $ & 
        \cellcolor{yellow}
        $E[Y_1 (0)-Y_1 (1)\mid G=1]-E[Y_0 (1)-Y_0 (0)\mid G=1] $ & 
        \cellcolor{yellow}
        $E[Y_0 (0)-Y_0 (1)\mid G=0]-E[Y_1 (1)-Y_0 (0)\mid G=0] $ & 
        \cellcolor{lightgreen}
        $E[Y_1 (1,0)-Y_1 (1,1)\mid G=1]-E[Y_1 (0,1)-Y_1 (0,0)\mid G=0]$ \\	
        \hline
        \textbf{h) Pre-post switch off} & 
        \cellcolor{yellow}
        $E[Y_0 (1)-Y_0 (0)\mid G=0]-E[Y_0 (1)-Y_0 (0)\mid G=1] $ (*also interesting w/ carryover) & 
        \cellcolor{lightgreen}
        $E[Y_1 (1,0)-Y_1 (1,1)\mid G=1]-E[Y_1 (1,0)-Y_1 (1,1)\mid G=0] $ & 
        \cellcolor{yellow}
        $(E[Y_1 (0)-Y_1 (1)\mid G=1]+E[Y_0 (0)-Y_0 (1)\mid G=1])- (E[Y_1 (0)-Y_1 (1)\mid G=0]+E[Y_0 (0)-Y_0 (1)\mid G=0]) $ & 
        \cellcolor{grey}
        $0$ & 
        \cellcolor{lightgreen}
        $E[Y_1 (1,0)-Y_1 (1,1)\mid G=1]- E[Y_1 (1,0)-Y_1 (1,1)\mid G=0] $ \\
        \hline
    \end{tabular}
    \caption{Causal contrasts identified by the gDiD formula for each combination of data structure from Figure \ref{fig:data_structures} (rows) and parallel trends assumption from Table \ref{tab:pts} (columns). Direct causal comparisons at a single time point and group are in dark green; potentially useful differences between effects at different times or groups are in light green; direct causal comparisons at a single time point and group that require a `no carryover' assumption are in orange; potentially useful differences between effects at different time points or groups that require a `no carryover' assumption are in yellow; and apparently not useful contrasts are in gray.}\label{tab:array}
\end{table}

Table \ref{tab:array} displays the causal contrast identified by gDiD for each combination of data structure (from Figure \ref{fig:data_structures}) and parallel trends assumption (from Table \ref{tab:pts}). We distinguish between contrasts that are average treatment effects in some subgroup at a particular time (dark green) and contrasts that are \emph{differences} between effects across subgroups or times (light green).
Sometimes an interesting causal contrast could only be obtained by imposing an additional `No Carryover' assumption:
\begin{equation}\label{eq:no_carryover}
Y_1(a_0,a_1) = Y_1(a_0',a_1) \mbox{ for all } a_0 \mbox{ and } a_0'\;,
\end{equation}
that is, that the time $0$ treatment had no impact on the time $1$ outcome. 
In these cases, we report the estimand derived by invoking this additional assumption, and we color the cell orange if it is an effect in a single group and time or yellow if it is a difference between effects at different groups or times. Some estimands we could not imagine being useful; these are in gray cells. 
In the following subsections, we discuss selected cells that illustrate the interplay between data structure and assumptions. 
Derivations for each cell are reported in Appendix~\ref{section_proofs}.

\subsection{The Canonical DiD setting: Panel (a), `Untreated' Parallel Trends}\label{canonical}
\begin{figure}[ht]
    \centering
\includegraphics[scale=1]{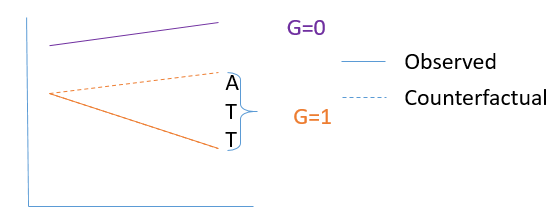}
    \caption{Schematic depicting the `untreated' parallel trends assumption in the canonical setting of panel (a) in Figure \ref{fig:data_structures}. The counterfactual trend in the treated group $G=1$ is assumed parallel to the observed untreated trend in the untreated group $G=0$.}
    \label{fig:cdid_pt}
\end{figure}

For completeness and for easy comparison to other settings, we present here the familiar identification derivation in the cDiD setting. In this case, we combine the `untreated' parallel trends assumption (first row of Table \ref{tab:pts}) with the data structure in which one group is treated only in the second period (panel a) of Figure~\ref{fig:data_structures}). 
Figure \ref{fig:cdid_pt} schematically illustrates the `untreated' parallel trends assumption in this setting. 
We can identify the ATT as follows:
\begin{align*}
     &E[Y_1-Y_0|G=1]-E[Y_1-Y_0|G=0]\\
    =&(E[Y_1-Y_0|G=1]-E[Y_1(0,0)-Y_0(0)|G=1])-\\
     &(E[Y_1-Y_0|G=0]-E[Y_1(0,0)-Y_0(0)|G=0])\\
    =&(E[Y_1(0,1)-Y_0(0)|G=1]-E[Y_1(0,0)-Y_0(0)|G=1])-\\
     &(E[Y_1(0,0)-Y_0(0)|G=0]-E[Y_1(0,0)-Y_0(0)|G=0])\\
    =&E[Y_1(0,1)-Y_1(0,0)|G=1]\;.
\end{align*}
Using the `untreated' parallel trends assumption, we subtract the changes in untreated potential outcomes of each group (which are assumed equal) from each term (first equality). 
Then we apply the Consistency assumption of Eq.~(\ref{eq:consistency}) to replace observed with potential outcomes for this data structure (second equality).
Then we cancel terms using linearity of expectation (third equality). 

The plausibility of this parallel trends assumption can be partially assessed by looking at trends in the outcome prior to treatment in the treated and untreated groups, as discussed in Section \ref{section_principles}.

\subsection{The Pre-Post Setting: Panel (b), `Untreated' Parallel Trends}\label{prepost}
Now consider the pre-post setting (panel (b) in Figure \ref{fig:data_structures}) and again invoke the `untreated' parallel trends assumption (row 1 of Table \ref{tab:pts}).
Eq.~(\ref{eq:gdid}) identifies the difference between the second period ATTs in the two groups, i.e.,
\begin{equation}\label{estimand_group}
E[Y_1(0,1)-Y_1(0,0)|G=1] - E[Y_1(0,1)-Y_1(0,0)|G=0].
\end{equation}
The derivation is similar to the canonical case above,
\begin{align}
\begin{split}\label{prepost_derivation}
    &E[Y_1-Y_0|G=1] - E[Y_1-Y_0|G=0]\\
=&(E[Y_1-Y_0|G=1]-E[Y_1(0,0)-Y_0(0)|G=1]) -\\
&(E[Y_1-Y_0|G=0]-E[Y_1(0,0)-Y_0(0)|G=0])\\
=&(E[Y_1(0,1)-Y_0(0)|G=1]-E[Y_1(0,0)-Y_0(0)|G=1]) - \\
&(E[Y_1(0,1)-Y_0(0)|G=0]-E[Y_1(0,0)-Y_0(0)|G=0])\\
=&E[Y_1(0,1)-Y_1(0,0)|G=1] - E[Y_1(0,1)-Y_1(0,0)|G=0].
\end{split}
\end{align}
Using the `untreated' parallel trends assumption, we subtract group-specific changes from both terms (first equality). 
Then we apply the Consistency assumption to replace observed with potential outcomes for this data structure: $A_0=0$ and $A_1=1$ for both groups (second equality). 
Then we cancel terms using linearity of expectation (third equality).

\begin{figure}
    \centering
    \includegraphics[scale=1]{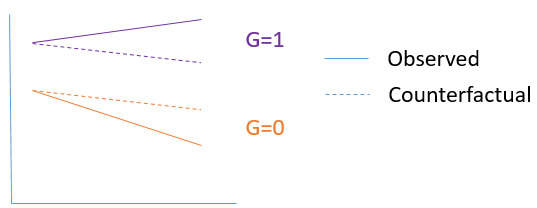}
    \caption{Illustration of the `untreated' parallel trends assumption in the pre-post setting of panel (b) from Figure \ref{fig:data_structures}. Unlike the cDiD setting, neither group's untreated trend is observed.}
    \label{fig:prepost_pt}
\end{figure}

The `untreated' parallel trends assumption in this setting is illustrated in Figure \ref{fig:prepost_pt}. 
What structural model might motivate using this causal assumption in this data setting?
Suppose untreated potential outcomes are generated by the structural model,
\begin{equation}\label{eq:structural_prepost}
    Y_t(\bar{0}_t) = \lambda_t + f(U,G) + \epsilon_t\;,
\end{equation}
where $U$ are time-invariant common causes of $G$ and the outcomes and we assume $E[\epsilon_t|U,G]=0$. 
This resembles the structural model of Eq.~(\ref{eq:fixed_effects}), but now $G$ represents a baseline variable that may cause untreated potential outcomes. 
Key implications of this structural model include:
\begin{itemize}
\item The joint contributions of $G$ and $U$ to the untreated potential outcomes are constant and additive via $f(G,U)$.
\item The time-specific shocks $\lambda_t$ affect both groups' untreated outcomes equally.
\item $U$ is sufficient to adjust for all confounding of $G$ and the outcomes (which follows from $E[\epsilon_t\mid U,G]=0$), and $Y_0$ is not a cause of $G$.
\begin{itemize}
  \item If $G$ is a time-invariant baseline covariate that precedes $Y_0$, this restriction is automatically satisfied.
  \item If $G$ is randomly assigned, like biological sex, $U$ will be empty. 
\end{itemize}
\end{itemize}
Therefore, parallel trends does not correspond to `additive equi-confounding' of $G$ and $Y_t(\bar{0})$ across time, because the association between $G$ and $Y_t(\bar{0})$ may contain both direct causation and common causes. 

\begin{figure}[ht]
  \centering
  \begin{subfigure}[b]{0.49\textwidth}
    \centering
    \includegraphics[width=\textwidth]{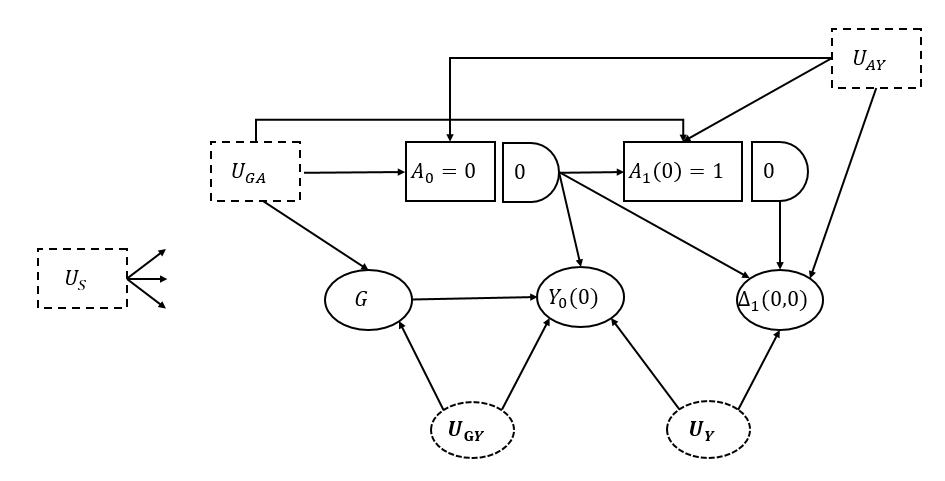}
    \caption{No $Y_0\rightarrow \Delta(0,0)$}
  \end{subfigure}
  \hfill
  \begin{subfigure}[b]{0.49\textwidth}
    \centering
    \includegraphics[width=\textwidth]{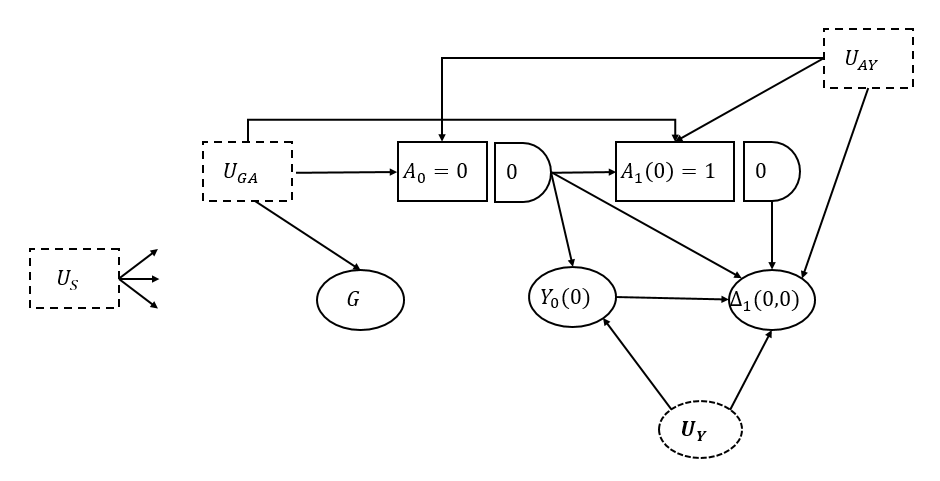}
    \caption{$G\independent Y_0$}
  \end{subfigure}
  \caption{SWIGs reflecting conditions under which the `untreated' parallel trends assumption holds in pre-post designs with shared treatment assignment across all units. Exposure is entirely determined by state- (i.e., population-) level shared variables, which are all implicitly conditioned on. The key constraints are no direct or backdoor paths between $G$ and $\Delta(0,0)$ at the unit (e.g., state resident) level. This requires no $G\rightarrow \Delta(0,0)$, no unit-level $G\leftarrow U \rightarrow \Delta(0,0)$ path, and either $G\independent Y_0$ or no $Y_0\rightarrow \Delta(0,0)$ arrow.}
  \label{fig:swigs_prepost_shared}
\end{figure}

\begin{figure}[ht]
  \centering
  \begin{subfigure}[b]{0.49\textwidth}
    \centering
    \includegraphics[width=\textwidth]{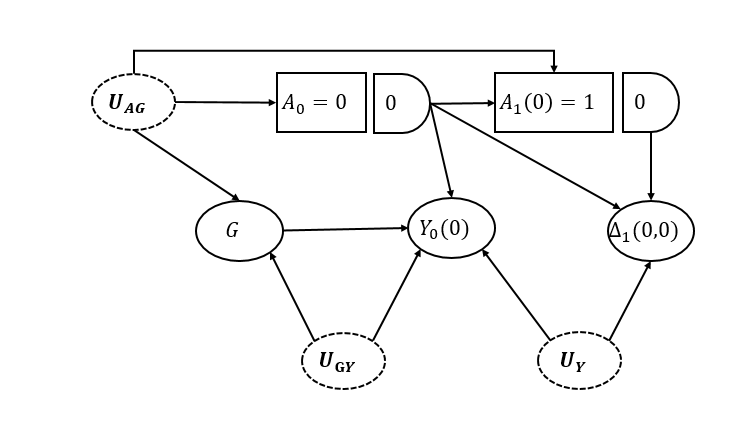}
    \caption{$A\independent \Delta(0,0)$}
  \end{subfigure}
  \hfill
  \begin{subfigure}[b]{0.49\textwidth}
    \centering
    \includegraphics[width=\textwidth]{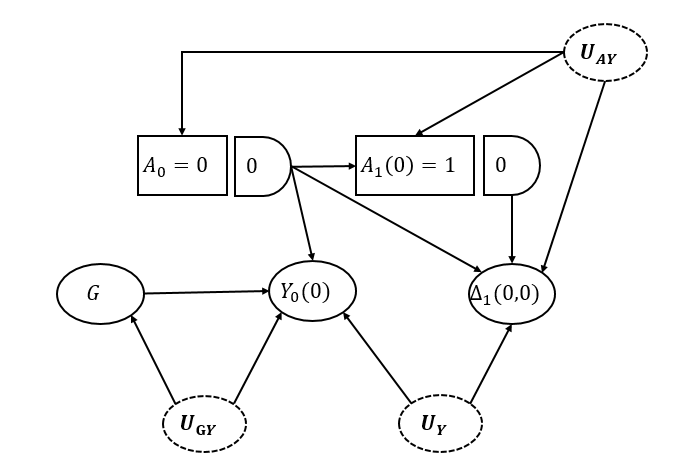}
    \caption{$G\independent A$}
  \end{subfigure}
  \caption{SWIGs reflecting conditions under which the `untreated' parallel trends assumption holds in pre-post designs with \textit{un}shared treatment assignment. This could mean exposure is assigned at the state level with person level units from multiple states, or it could mean that exposure is assigned at the same level as units (e.g. exposure assigned at the state level and units are states). 
  }
  \label{fig:swigs_prepost_unshared}
\end{figure}

Now, we use graphical reasoning to consider the `untreated' parallel trends assumption in pre-post applications. 
We distinguish between two settings in which this design might be applied, corresponding to the SWIGs in Figures \ref{fig:swigs_prepost_shared} and \ref{fig:swigs_prepost_unshared}. 
In the first setting, treatment is applied to the entire population of interest, i.e., all units share a treatment assignment mechanism. 
COVID is one such example, as no country could avoid being exposed to the pandemic, and we may wish to study differences in responses to this common exposure across different types of countries.
Another example is a policy implemented by a single state in which we wish to study heterogeneous responses among individuals with different characteristics, e.g., effects of Kansas's Medicaid expansion in rural versus urban residents of that state. 

In the second scenario, units have different treatment assignment mechanisms. 
One example is a study of a policy implemented in several states in which we wish to study heterogeneous effects among urban and rural residents of all the treated states. 
However, treatment need not be assigned at a `higher' level.
Another example is a study of Medicaid expansion in which we study heterogeneous effects across rural versus urban \textit{states}.
Below, we explain why this distinction is important and argue that the `untreated' parallel trends assumption in the pre-post setting is more plausible when treatment assignment is shared.

Recall, the `untreated' parallel trends assumption in this setting, 
$E[Y_1(0,0)-Y_0(0)|G=1] = E[Y_1(0,0)-Y_0(0)|G=0]$, has a somewhat different interpretation than the symbolically identical assumption in the canonical setting.
Here, we condition on groups defined by $G$, across which we wish to study differences in effects (e.g., rural vs urban in the examples above), rather than on the exposure.
If $G$ is independent of $\Delta(0,0)$, this parallel trends assumption will hold.
Again, we note that full independence is stronger than strictly required, but still useful to guide our thinking.

We consider SWIGs in which we intervene on treatment variables to consider potential outcomes under an `untreated' regime \emph{and} condition on both treatment variables, since in this design all units go from untreated to treated in the observed data.

Figure \ref{fig:swigs_prepost_shared} displays such SWIGs for the setting in which the treatment assignment mechanism is shared (e.g., all countries are exposed to COVID or a single state implements a policy).
However, the outcomes ($Y_0$ and $\Delta(0,0)$) and group membership ($G$) are not shared (e.g., they vary across multiple countries all exposed to COVID or across individuals in a single state).
We assume that these lower-level attributes would have negligible effects on the higher-level joint treatment assignment and hence omit arrows from $G$ and $Y_0$ into the treatment nodes. 

Population-level (i.e., shared) common causes of treatment and other variables are denoted by $U_{A*}$ nodes.
The $U_S$ variable with arrows emanating in all directions is meant to convey, while limiting visual clutter, that shared population-level variables may be causes of all variables in the graph.  
All these population-level shared variables--- the intervention and common causes of the intervention and population-level summaries of individual-level variables---are implicitly conditioned on (i.e., their value is identical for all individuals in the population by definition of `shared'). 

For the independence version of parallel trends to hold, we require no directed path ($G\rightarrow \Delta(0,0)$ or $G\rightarrow Y_0 \rightarrow \Delta(0,0)$) between $G$ and $\Delta(0,0)$ and no \textit{unit-level} common cause ($G\leftarrow U_{G\Delta}\rightarrow \Delta(0,0)$ or $G\leftarrow U_{GY}\rightarrow Y_0\rightarrow \Delta(0,0)$) of $G$ and $\Delta(0,0)$. 
(Population-level common causes of $G$ and $\Delta(0,0)$, captured by $U_S$ in the graph, are allowed, as they are implicitly conditioned on.) 
These assumptions might be violated, for example, if $G$ has a time-varying direct effect on the outcome or if $G$ is an effect modifier of another exposure that occurs at the same time as the intervention. 
If $G$ is associated with $Y_0$ within the population (i.e., one or both of $G\rightarrow Y_0$ and $G\leftarrow U_{GY}\rightarrow Y_0$ are present) as in panel (a), then there can be no $Y_0\rightarrow \Delta(0,0)$ arrow, like canonical DiD. 
If $G$ is independent of $Y_0$ in the population as in panel (b), then a $Y_0\rightarrow \Delta(0,0)$ arrow is permissible under parallel trends. While marginal independence of $G$ and $Y_0$ is empirically testable with unit level data, we expect that it would rarely be satisfied.

Suppose the groups of interest are rural and urban residents of a single state, and the exposure is a state-level policy. 
Collider paths running through the exposure, e.g., $G\leftarrow U_{AG} \rightarrow A_1 \leftarrow U_{AY} \rightarrow \Delta(0,0)$, are all blocked because all causes of the exposure are state-level and therefore implicitly conditioned on. 
Unlike canonical DiD, state-level confounders of the exposure and trend \textit{across states} may be present without leading to violations.
We can also allow aggregate state-level baseline outcomes (e.g., proportion uninsured in the Medicaid example) to be a cause of the exposure $A_1$, again in contrast to the canonical setting. 
Thus, `untreated' parallel trends in a pre-post setting within a state where exposure is shared across units may hold in many cases where the canonical DiD `untreated' parallel trends assumption \emph{across} states fails. 

\clearpage

When treatment is \textit{not} shared across units in the population, the causes of treatment are no longer implicitly conditioned on. 
Thus, there can no longer be state-level common causes of $G$ and $\Delta(0,0)$, reflected by the absence of a $U_S$ node in the SWIGs of Figure \ref{fig:swigs_prepost_unshared}.
Further, collider paths through $A_1=1$, which is still conditioned on in a pre-post design, are now potentially open. 
To block backdoor paths from $G$ to $\Delta(0,0)$ via $A_1$, we must assume that $A$ is independent of $G$ (i.e., no $G\rightarrow A$ or $G \leftarrow U_{GA}\rightarrow A$ paths) as in Figure \ref{fig:swigs_prepost_unshared} panel (b) or that $A$ and $\Delta(0,0)$ are independent as in Figure \ref{fig:swigs_prepost_unshared} panel (a) (which implies the same graphical restrictions as cDiD parallel trends as well as additional restrictions on the association between $G$ and $\Delta(0,0)$). 

Independence of $G$ and $A$ is empirically checkable with access to a full data set including untreated states. 
Recall, when treatment is not shared across units, treatment may be assigned at the unit level (e.g. if units are states, and the sample comprises states that independently implemented a policy of interest). 
In that case, $G\rightarrow A$ or $Y_0\rightarrow A_1$ arrows are possible.
The absence of both $G\rightarrow A$ and $Y_0\rightarrow A$ arrows in Figure \ref{fig:swigs_prepost_unshared} panel (b) and $Y_0\rightarrow A_1$ in Figure \ref{fig:swigs_prepost_unshared} panel (a), required to avoid open collider paths between $G$ and $\Delta(0,0)$, are then strong and meaningful restrictions. 
In the case that units are individuals, treatment is assigned at the state level, and the population comprises individuals from multiple states, then one would expect (as in the shared exposure scenario) that there are no $G\rightarrow A$ or $Y_0\rightarrow A_1$ arrows, but collider paths via unobserved state level variables remain a concern.

Returning to our running example, if we interpret COVID-19 as the treatment and UHC as the baseline covariate defining groups, then this study has a pre-post data structure.
If we invoke the untreated parallel trends assumption, we can summarize the findings of \citet{kim2024synergistic} as follows: the onset of the COVID-19 pandemic reduced immunization coverage by 1.14 fewer percentage points (95\% CI: 0.39\%, 1.90\%) in high UHC countries compared to low UHC countries. 
Notice that this interpretation does not ascribe a causal role to UHC in modifying the effect of COVID-19. 
Even under our identifying assumptions, UHC may simply be associated with varying effects of COVID-19 without being responsible for that variation \citep{vanderweele2007four}.

Moreover, because under the structural model in Eq.~(\ref{eq:structural_prepost}), `untreated' parallel trends holds at all time points, we might examine baseline trend differences between groups defined by $G$.
Indeed, \citet{kim2024synergistic} found that average national vaccination rates in high and low UHC countries fluctuated similarly prior to COVID-19, and offered this as evidence for the plausibility of the `untreated' parallel trends assumption, ``For all analyses, we checked whether the parallel pre-trend assumption was satisfied."

The estimand (\ref{estimand_group}) might generally be useful for studying whether effects of a policy were equitable. For example, did Medicaid expansion have similar effects across racial groups \citep{moriya2023racial}? If effect heterogeneity is of primary interest, then an investigator might choose to compute a gDiD estimate in only the treated even if a control group is available, particularly if exposure assignment is shared across the population. If it is within the investigator's control, the investigator may wish to limit the sample to units that do share an exposure assignment mechanism, e.g., residents of a single state, or perform separate analyses within populations that each share exposure assignments, so that weaker substantive assumptions are required due to implicit conditioning as discussed above. The conditional parallel trends assumption within levels of $G$ across treatment groups \citep{abadie2005semiparametric} required by a standard approach might be less plausible than the group `untreated' parallel trends assumption in the treated. For example, as we saw in our graphical analyses, if (aggregate) $Y_0$ causes $A_1$ at the state level, then canonical parallel trends will not hold. However, `untreated' parallel trends in a pre-post design might. 

\subsection{The No Pre-Period Setting: Panel (c), `Untreated' Parallel Trends}\label{nopre}
In the no pre-period setting (panel (c) of Figure~\ref{fig:data_structures}) with the `untreated' parallel trends assumption (row 1 of Table \ref{tab:pts}), 
the gDiD formula in Eq. (\ref{eq:gdid}) identifies the difference between the ATTs in the second vs first time periods, i.e.,
\begin{equation}\label{estimand_nopre}
E[Y_1(1,1)-Y_1(0,0)|G=A_0=A_1=1] - E[Y_0(1)-Y_0(0)|G=A_0=A_1=1].
\end{equation}
The identification derivation is similar to the cDiD and pre-post cases above, 
\begin{align}
\begin{split}\label{nopre_derivation}
    &E[Y_1-Y_0|G=1] - E[Y_1-Y_0|G=0]\\
=&(E[Y_1-Y_0|G=1]-E[Y_1(0,0)-Y_0(0)|G=1]) -\\
&(E[Y_1-Y_0|G=0]-E[Y_1(0,0)-Y_0(0)|G=0])\\
=&(E[Y_1(1,1)-Y_0(1)|G=1]-E[Y_1(0,0)-Y_0(0)|G=1]) - \\
&(E[Y_1(0,0)-Y_0(0)|G=0]-E[Y_1(0,0)-Y_0(0)|G=0])\\
=&E[Y_1(1,1)-Y_1(0,0)|G=1] - E[Y_0(1)-Y_0(0)|G=1]\\
\end{split}
\end{align}
As in the previous derivations, the parallel trends assumption allows us to subtract the same quantity from both terms (first equality). 
Then we apply Consistency; in this setting $A_0=A_1=G$ for all units (second equality).
Finally, we cancel terms (third equality).

\begin{figure}[h]
    \centering
    \includegraphics[scale=1]{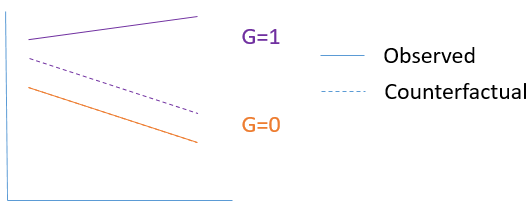}
    \caption{Illustration of the `untreated' parallel trends assumption in the no pre-period setting of panel (c) from Figure \ref{fig:data_structures}. Like the cDiD setting, the untreated trend in the $G=0$ group is observed. Unlike both the cDiD and pre-post settings, we do not get to observe the untreated potential outcome in the $G=1$ group at either time point.}
    \label{fig:nopre_pt}
\end{figure}

The `untreated' parallel trends assumption in this setting is illustrated in Figure \ref{fig:nopre_pt}. Unlike the cDiD and pre-post settings, the plausibility of the group parallel trends assumption Eq.~(\ref{gpt}) cannot be assessed via pre-trends in this setting because we do not get to observe untreated outcomes in the treated group (unless the treated group was untreated prior to baseline, in which case a canonical DiD analysis focusing on the period when treatment first began might be more desirable). 
Furthermore, assuming that the treated group was treated prior to time $0$, the structural model Eq.~(\ref{eq:fixed_effects}) would not necessarily apply to the counterfactual outcomes under no treatment only at times $0$ and $1$, as prior treatment might have `changed the fundamentals' over the study period. 

Returning to our running example, if we consider UHC the treatment, the study has the `no pre-period' data structure [panel (c) of Figure~\ref{fig:data_structures}].
Invoking the `untreated' parallel trends assumptions, we could summarize their results as follows: high UHC increased immunization coverage by 1.14 additional percentage points (95\% CI: 0.39, 1.90) in the post-COVID period compared to the pre-COVID period. 
Again, this phrasing does not ascribe a causal role to the pandemic in modifying the effect of high UHC over time. 
However, this is unlikely to be the authors' intended interpretation, because this data structure does not allow us to look for parallel trends in untreated outcomes. 
The fact that \citet{kim2024synergistic} performed tests for parallel trends in the pre-COVID period suggests that they were not targeting the estimand in Eq.~(\ref{estimand_nopre}) with UHC as the treatment. 

Next, we consider some novel parallel trends assumptions and their combinations with data structures that yield useful estimands.

\subsection{`Switch On' vs. Always Treated: Panel (d), `Never Switch' Parallel Trends}\label{section_switch_on_vs_always}
\begin{figure}
    \centering
    \includegraphics[scale=1]{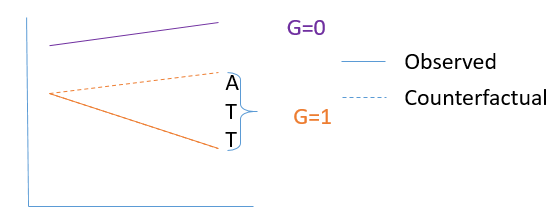}
    \caption{Illustration of the `never switch' parallel trends assumption in the `switch on vs always treated' setting of panel (d) from Figure \ref{fig:data_structures}. The illustration is identical to the cDiD case! The only difference is in the observed and counterfactual regimes associated with the two groups.}
    \label{fig:switchon_ns_pt}
\end{figure}

Panel (d) of Figure \ref{fig:data_structures} depicts a scenario in which treatment `switches on' in one group, while it is always present in the other. In this setting, the array in Table \ref{tab:array} shows that the usual `untreated' parallel trends assumption does not identify a meaningful estimand. 
However, under the `never switch' parallel trends assumption from row 5 of Table \ref{tab:pts}, the gDiD formula in Eq.~(\ref{eq:gdid}) identifies
\begin{equation}
    E[Y(0,1)-Y(0,0)|G=1],
\end{equation}
i.e., the effect of initiating treatment in the second time point compared to never initiating treatment (in the group that did initiate at the second time point).
The derivation is similar to previous subsections and can be found in Appendix~\ref{section_proofs}. The `never switch' parallel trends assumption in this setting is illustrated in Figure \ref{fig:switchon_ns_pt}

If units had maintained their time $t=0$ treatment status for an extended period prior to $t=0$, it may be reasonable to partially assess the `never switch' parallel trends assumption using pre-trends. 
Note that just as the `untreated' parallel trends assumption could be rearranged to yield `additive equi-confounding', the `never switch' assumption in this data structure can be rearranged to yield:
\begin{equation}\label{additive_equi-association_never_treat}
    E[Y_1(\bar{1})|G=1]-E[Y_1(\bar{0})|G=0]=E[Y_0(\bar{1})|G=1]-E[Y_0(\bar{0})|G=0].
\end{equation}
This states that the combined additive association generated by the effect of sustained treatment at the observed baseline level and by confounding is constant across time. 
Let $A_{base}$ denote the observed baseline level of treatment. 
One might capture this additive equi-association dynamic in a structural model,
\begin{align}
\begin{split}\label{structural_never_switch}
Y_t(\bar{A}_{base})=\lambda_t+f(A_{base},U)+\epsilon_{t}\;,
\end{split}
\end{align}
where $Y_t(\bar{A}_{base})$ denotes the outcome under continual treatment at observed baseline treatment $A_{base}\in\{0,1\}$ and $U$ comprises all confounders of $A_{base}$ and the outcome. In this model, 
\begin{itemize}
\item the joint effect of confounders and cumulative treatment at the observed baseline value does not depend on time. 
\item All time-specific `shocks' have similar average effects in the (baseline) treated and untreated groups. 
\end{itemize}
The assumption that cumulative treatment effects do not vary with time might be reasonable under scenarios discussed in Section \ref{section_treatment_fundamentals}, in which the evolution of the treated units has returned to an equilibrium state. 
Under the consistency assumption, this structural model implies parallel pre-trends in the observed outcomes and can thus be tested empirically. 

\begin{figure}[ht]
    \centering
    \includegraphics[scale=0.8]{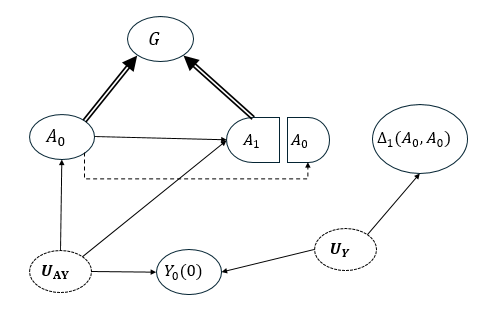}
    \caption{A Single World Intervention Graph illustrating implications of the `never switch' parallel trends assumption.}
    \label{fig:never_switch_swig}
\end{figure}

We represent the `never switch' parallel trends assumption graphically in a SWIG in Figure \ref{fig:never_switch_swig}. $\Delta(A_0,A_0)$ represents the counterfactual trend under the `never switch' regime. The double arrows from $A_0$ and $A_1$ into $G$ indicate that $G$ is a deterministic function of $A_0$ and $A_1$ \citep{richardson2013single}. The dotted arrow from $A_0$ to the intervention node setting $A_1$ to $A_0$ represents a dynamic treatment regime in which the treatment intervention is a function of past observed treatment. For $A_0$ and $A_1$ (and therefore $G$) to be independent of $\Delta(A_0,A_0)$, there cannot be a $Y_0\rightarrow A_1$ arrow, a $Y_0\rightarrow \Delta(A_0,A_0)$ arrow, or a backdoor $A\leftarrow U \rightarrow \Delta(A_0,A_0)$ path, just like in the SWIG in Figure \ref{fig:cdid_swig} representing the cDiD parallel trends assumption. Furthermore, there can be no effect of $A_0$ on $\Delta(A_0,A_0)$, reflected in the absence of arrows into $\Delta(A_0,A_0)$ from either the observed $A_0$ or the intervention node at time 1 setting $A_1$ to $A_0$. This is consistent with the requirement that any effects of $A_0$ are constant over time and thus cancel out of the difference in $\Delta(A_0,A_0)$ at the chosen scale. 

Note that if $A_0$ is constant across units, then we condition on it and `never switch' reduces to `untreated' or `treated' parallel trends. Because $A_0$ is conditioned on, arrows from the observed time $0$ treatment  and from the time $1$ intervention node to $\Delta(A_0,A_0)$ become permissible, as in the canonical setting. It is only when $A_0$ differs between groups that the structural requirements for `never switch' become somewhat more restrictive than those for canonical parallel trends.

\subsection{`Switch Off' vs `Untreated': Panel (f)}
Here, the `never switch' and `untreated' parallel trends assumptions lead to identification of different potentially useful estimands by the gDiD formula. Under the `never switch' parallel trends assumption, the gDiD formula identifies
\begin{equation}
    E[Y_1(1,0)-Y_1(1,1)|G=1],
\end{equation}
i.e., the effect of switching off treatment compared to remaining treated at the second time point in units that switched off treatment. This estimand directly assesses whether stopping treatment was beneficial in those that stopped. 
Under the `untreated' parallel trends assumption, the gDiD formula identifies
\begin{equation}
    E[Y_1(1,0)-Y_1(0,0)|G=1]-E[Y_0(1)-Y_0(0)|G=1].
\end{equation}
This is the difference between the lasting effect of treatment at the first time point on the outcome at the second time point and the immediate effect of treatment at the first time point. This estimand answers the question: ``How quickly does the effect of treatment dissipate after it is removed?" While the estimand identified under `never switch' seems more practical, the gDiD formula should be interpreted based on which assumption is more plausible.

\begin{figure}
    \centering
    \includegraphics[scale=1]{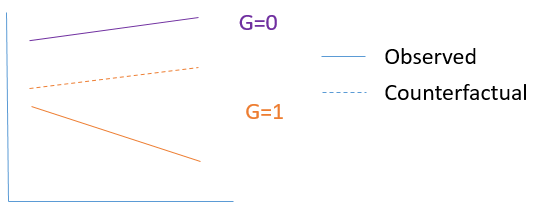}
    \caption{Illustration of the `untreated' parallel trends assumption in the `switch off vs untreated' setting of panel (f) from Figure \ref{fig:data_structures}. Like the no pre-period setting, the untreated trend in the $G=0$ group is observed, and we do not get to observe the untreated potential outcome in the $G=1$ group at either time point.}
    \label{fig:switchoff_untreated_pt}
\end{figure}

\begin{figure}
    \centering
    \includegraphics[scale=1]{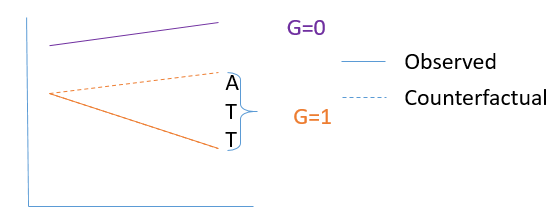}
    \caption{Illustration of the `never switch' parallel trends assumption in the `switch off vs untreated' setting of panel (f) from Figure \ref{fig:data_structures}. Again, the visual depiction is identical to the cDiD case, but with different observed and counterfactual regimes corresponding to the two groups.}
    \label{fig:switchoff_ns_pt}
\end{figure}

Assuming units received their baseline treatment assignments prior to time $0$, only the `never switch' assumption admits pre-trends tests. This is because we do not get to observe untreated outcomes prior to time $0$ in the `switch off' group. Furthermore, lasting effects of prior treatment might invalidate the `untreated' parallel trends assumption over the study period. Supposing the `never switch' pre-trends test fails, it would seem to require a bit a of a leap of faith to adopt the `untreated' parallel trends assumption. Illustrations of the two assumptions in this setting are provided in Figures \ref{fig:switchoff_untreated_pt} and \ref{fig:switchoff_ns_pt}.

If the `never switch' assumption holds, might the `untreated' assumption hold as well? If both assumptions held, then the two causal estimands stated above would be equal. An algebraic rearrangement of this equality is that
\begin{equation}
    E[Y_1(1,1)-Y_1(0,0)|G=1] = E[Y_0(1)-Y_0(0)|G=1],
\end{equation}
i.e., that the effect of sustained treatment is equal to the immediate effect of treatment in the group that is treated at baseline. (We can also obtain this constraint by differencing the two assumptions.) As we discussed in Section \ref{section_switch_on_vs_always}, this constraint is already encoded in the structural model (\ref{structural_never_switch}) that we put forth to justify the `never switch' parallel trends assumption. Thus, combining these two particular parallel trends assumptions in this setting does not result in additional unpalatable implications regarding effect homogeneity.

\section{Extensions}\label{section_extensions}
We discuss some straightforward extensions in the setting of a pre-post data structure under the `untreated' parallel trends assumption. We focus on this setting due to space constraints and because it is an interesting non-canonical setting, but similar extensions would apply to other settings. 

\subsection{Conditional group parallel trends}
Sometimes a parallel trends assumption may be more plausible within levels of other covariates $Z$. 
For instance, we might believe a parallel trends assumption on consumption holds conditional on household income.
Suppose we make the conditional `untreated' parallel trends assumption,
\begin{equation}
    E[Y_1(0)-Y_0|G=1,Z]=E[Y_1(0)-Y_0|G=0,Z].
\end{equation}
In the pre-post setting, we can adapt Abadie's \citep{abadie2005semiparametric} inverse probability weighted DiD identification formula for the ATT to an inverse probability weighted gDiD identification formula for effect modification of the ATT:
\begin{equation}
   E[Y_1(1)-Y_1(0)|G=1] -  E[Y_1(1)-Y_1(0)|G=0]=E\left[\frac{Y_1-Y_0}{P(G=1)}\frac{G-P(G=1|Z)}{1-P(G=1|Z)}\right].
\end{equation}
The proof exactly follows the proof of Lemma 3.1 of \citep{abadie2005semiparametric}. A corresponding plug-in estimator is given by
\begin{equation*}
    \mathbb{P}_n\left\{\frac{Y_1-Y_0}{\hat{P}(G=1)}\frac{G-\hat{P}(G=1|Z)}{1-\hat{P}(G=1|Z)}\right\}
\end{equation*}
where $\mathbb{P}_n$ denotes sample average, $\hat{P}(G=1)$ denotes the sample proportion of $G=1$, and $\hat{P}(G=1|Z)$ denotes an estimate (e.g., via logistic regression) of the conditional probability that $G=1$ given covariate(s) $Z$.

\subsection{Continuous or multi-valued G}
Suppose $G$ is a continuous variable. In the pre-post setting, $G$ might be a continuous baseline covariate such as income. For two values $g$ and $g'$ of $G$, we can make the group parallel trends assumption 
\begin{equation}\label{continuous_gpt}
    E[Y_1(0)-Y_0(0)|G=g]=E[Y_1(0)-Y_0(0)|G=g'].
\end{equation}
Substituting $g$ for 1 and $g'$ for 0 in the derivations from the previous sections shows that under (\ref{continuous_gpt}) the gDiD expression $E[Y_1-Y_0|G=g]-E[Y_1-Y_0|G=g']$ identifies $E[Y_1(1)-Y_1(0)|G=g]-E[Y_1(1)-Y_1(0)|G=g']$ in the pre-post setting. Note that separate group parallel trends assumptions are required for any two levels of $G$ to be compared. Furthermore, a plugin gDiD estimator 
\begin{equation*}
 \hat{E}[Y_1-Y_0|G=g]-\hat{E}[Y_1-Y_0|G=g'],
 \end{equation*}
 where $\hat{E}[\cdot | \cdot]$ denotes estimated conditional expectations, 
would require a regression model for $E[Y_1-Y_0|G]$ in the continuous case where sample averages would suffice in the binary setting. 
\subsection{Bounds (or point identification with a zero effect subgroup)}
Note that if, as in triple differences, there exists a subgroup in which treatment is known to have no effect, gDiD might identify conditional effects in other subgroups. 
That is, if group parallel trends Eq.~(\ref{gpt}) holds with subgroups $G=1$ and $G=0$ and the intervention is known not to have any effect in subgroup $G=0$, then the second term in Eq.~(\ref{estimand_group}) is equal to 0 and the gDiD expression identifies
the ATT in subgroup $G=1$ (i.e., $E[Y_1(1)-Y_1(0)|G=1]$) in the pre-post setting. 
The zero effect subgroup is essentially an untreated control group and gDiD reduces to canonical DiD with a control group.

We can generalize this reasoning to get bounds on subgroup effects on the treated. 
If we assume that $\tau_l<E[Y_1(1)-Y_1(0)|G=0]<\tau_u$, then we have that 
\begin{equation*}
   \Delta\Delta + \tau_l < E[Y_1(1)-Y_1(0)|G=1] < \Delta\Delta +\tau_u,
\end{equation*}
where $\Delta\Delta = E[Y_1-Y_0|G=1] - E[Y_1-Y_0|G=0]$.

\section{Illustrative Application: Medicaid Expansion and Uninsurance}\label{application}
In this application, we explore alternative approaches to estimating the disparate impact of Medicaid expansion under the Affordable Care Act on uninsurance levels in non-Hispanic Black and White people. 
The first approach we consider is the to take the difference between separate canonical DiD estimates in the two race groups. 
For this group specific approach, we invoke separate canonical parallel trends assumptions within each race group. 
The second approach applies the pre-post gDiD formula to the expanders, with $G$ denoting race. 
For pre-post gDiD, we invoke the pre-post `untreated' parallel trends assumption. 
We do not consider uncertainty quantification in this illustration, only point estimates. 
We are ultimately able to reason from a combination of empirical effect estimates and SWIGs that the `untreated' parallel trends assumption does not hold for either data structure. 

In a canonical DiD analysis, our data (panel (a) of Figure $\ref{fig:data_structures}$) comprised $2013$ ($t=0$) and $2014$ ($t=1$) uninsurance rates in states that expanded ($G=1$) and did not expand ($G=0$) in 2014, which was the first year that expansion was an option. 
We estimated that expansion in 2014 reduced the uninsurance rate by $0.88$ percentage points (pp) in the full Black and White population of 2014 expansion states. 
In subgroup analyses, we estimated that expansion reduced uninsurance by $0.95$ percentage points (pp) in the non-Hispanic White group and $1.1$ pp in the non-Hispanic Black group. 
Thus, the group-specific canonical approach yields an estimated $0.15$ pp additional reduction in the Black group. 

We could not find good reason to reject the underlying `untreated' parallel trends assumption on structural grounds (Figure \ref{fig:cdid_swig}). 
Because the decision to expand Medicaid was almost entirely driven by partisan control of individual state governments, any effect of prior uninsurance on expansion ($Y_0\rightarrow A_1$) was likely negligible. 
We could also not think of plausible time-varying factors that impact both state political lean and uninsurance rates. 
It is more difficult to reason about whether a $Y_0\rightarrow\Delta_1(0,0)$ arrow is present, since it depends on the specific factors that impact pre-treatment uninsurance rates. 
See Section 5 of \citet{renson2025using} for further discussion. 
Inability to reject parallel trends on structural grounds, of course, should not be interpreted as reason to accept it. 

Similarly, pre-trends plots (Figure \ref{uninsurance_trends}) do not provide evidence against the `untreated' parallel trends assumption in either subgroup. 
Differential differences in the pre-periods do not exceed approximately half the effect estimate in either group. 
However, the magnitudes of these violations do not provide reassurance that the estimate of the difference between the effects in the two groups is approximately correct. 
Again, failure to reject does not imply that we should accept. 

Next, we estimated the difference between effects in Black ($G=1$) and White ($G=0$) groups from a pre-post data structure (panel (b) of Figure $\ref{fig:data_structures}$) containing rates in 2013 ($t=0$) and 2014 ($t=1$) in the 2014 expansion states under an `untreated' parallel trends assumption with the gDiD formula:
\begin{equation*}
    (Rate^{2014}_{Black} - Rate^{2013}_{Black})-(Rate^{2014}_{White} - Rate^{2013}_{White}).
\end{equation*}
We estimated that expansion reduced uninsurance rates by $1.6$ pp more in Black than White people in expansion states. 
This difference in effects across groups exceeds the total estimated effect for Black people from the canonical DiD approach. 
How can we reconcile these results?

Pre-trends plots (Figure \ref{uninsurance_trends}) suggest that uninsurance rates in the two race/ethnicity groups changed similarly prior to expansion in expansion states. The maximum pre-exposure deviation in trends was less than half the $1.6$ pp estimate, and deviations decreased closer to the time of the exposure. 
Thus, pre-trends plots do not provide reason to reject the pre-post `untreated' parallel trends assumption. However, in the non-expanding states (another `placebo', like pre-treatment periods) the decrease in uninsurance rate was $1.5$ pp greater in Black people ($3.5$ pp) than White people ($2.1$ pp). This deviation from parallel trends in the untreated group at the year of treatment is good empirical reason to doubt the pre-post`untreated' parallel trends assumption. 

What does structural reasoning tell us? 
Treatment, in this setup, is unshared because we pooled residents of all states that independently decided to expand Medicaid. 
Therefore, we are in the setting of Figure \ref{fig:swigs_prepost_unshared}. 
In particular, because $G$ is not independent of $A$ (proportion Black was $0.09$ in expanding states and $0.15$ in non-expanding states), if pre-post untreated parallel trends holds, then we are in panel (a) of Figure \ref{fig:swigs_prepost_unshared}. 
This setting requires that $A_1\independent\Delta_1(0,0)$, which is precisely what is required by the marginal (i.e., not subgroup-specific) canonical parallel trends assumption. 
Thus, in this application, the pre-post untreated parallel trends assumption is strictly stronger than the marginal canonical parallel trends assumption. 
This means that if only one of the pre-post and marginal canonical estimates is correct, it is the marginal canonical one. 
And furthermore, if both the marginal canonical and pre-post untreated parallel trends assumptions hold, then the canonical parallel trends assumption within levels of $G$ (race) also holds (i.e., conditioning on $G$ in Figure \ref{fig:swigs_prepost_unshared} (a) does not open any violating paths)
However, this is an empirical contradiction because the pre-post and group-specific canonical estimates disagree. 
Therefore, either a) only the marginal canonical estimate is correct or b) neither the marginal canonical nor pre-post estimates are correct. 

Suppose the marginal canonical parallel trends assumption holds (and therefore the pre-post untreated parallel trends assumption does not hold). 
Is it possible that the conditional on race canonical parallel trends assumptions hold, too? 
Failure of pre-post untreated parallel trends implies the presence of an open path between $G$ (race) and $\Delta_1(0,0)$. 
It must be a backdoor path of the form $G\leftarrow\rightarrow \Delta_1(0,0)$, since $A_1\leftarrow U_{AG}\rightarrow G\rightarrow\Delta_1(0,0)$ would violate the marginal canonical assumption that we are assuming holds. 
Conditioning on the collider $G$, then, would open the path $A_1\leftarrow U_{AG} \rightarrow G \leftarrow\rightarrow\Delta_1(0,0)$, violating the conditional canonical parallel trends assumptions needed for validity of the group-specific canonical approach.
We have thus shown that if the marginal canonical assumption holds, then the conditional canonical assumption does not.

Conversely, it is technically possible that the conditional canonical assumption holds while the marginal does not. 
This would happen if the sole backdoor path violating the marginal conditional assumption were $A_1\leftarrow U_{AG} \rightarrow G \rightarrow\Delta_1(0,0)$. 
However, it would be quite coincidental if the effect modifier of interest ($G$, race) also perfectly filled this adjustment role.

To summarize, we conclude that, barring coincidence, either only the marginal canonical assumption or none of the assumptions holds in this case, meaning that we are unable to identify the difference between effects in Black and White groups via either the group-specific canonical or pre-post approach. 
Formally reasoning about the relative direction and magnitude of bias from each approach, while crucial, is for now out of scope.
(We note that these rates are not restricted to a low-income population likely to be Medicaid eligible under expansion. Many in the study population would have benefited from other concurrently introduced elements of the ACA, such as introduction of insurance exchanges, which likely explains why uninsurance rates also decreased substantially in the non-expansion groups in the treatment period.) 

\begin{figure}[ht]
  \centering
  \begin{subfigure}[b]{0.25\textwidth}
    \centering
    \includegraphics[scale=.4]{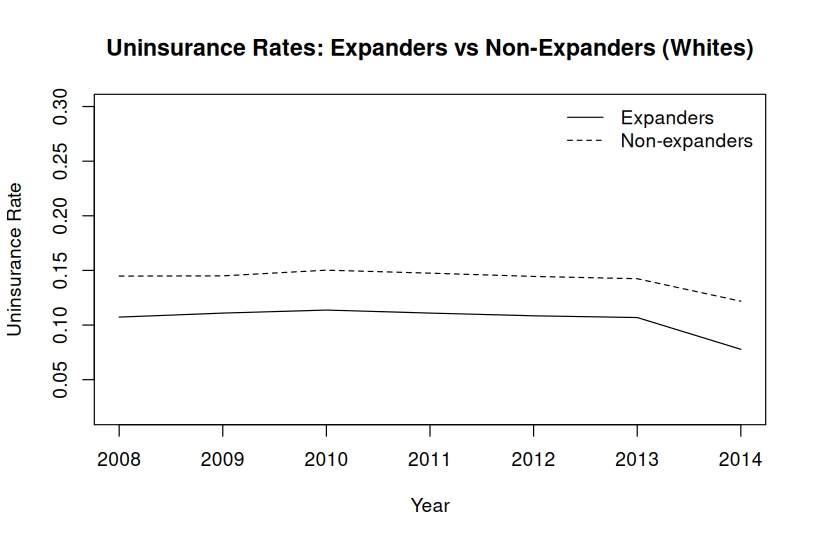}
    \caption{}
  \end{subfigure}\\
  \begin{subfigure}[b]{0.25\textwidth}
    \centering
    \includegraphics[scale=.4]{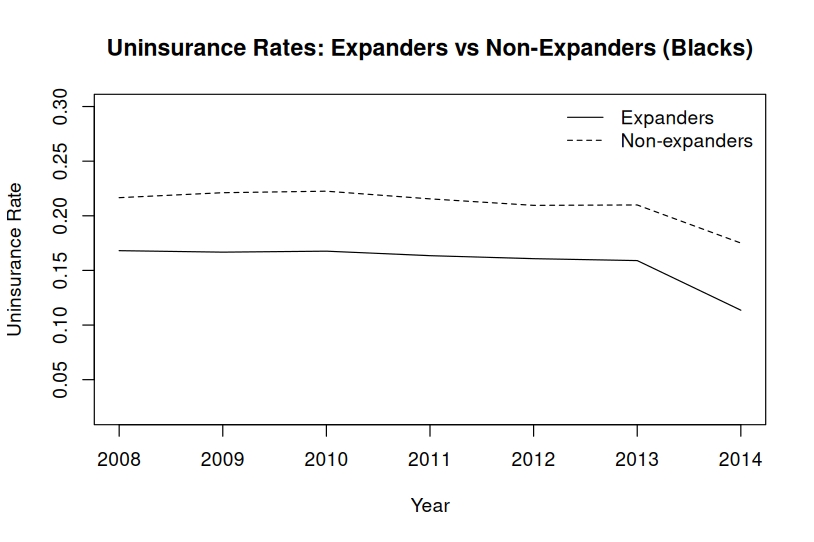}
   \label{fig:subfig2}
   \caption{}
  \end{subfigure}\\
  \begin{subfigure}[b]{0.25\textwidth}
    \centering
\includegraphics[scale=.4]{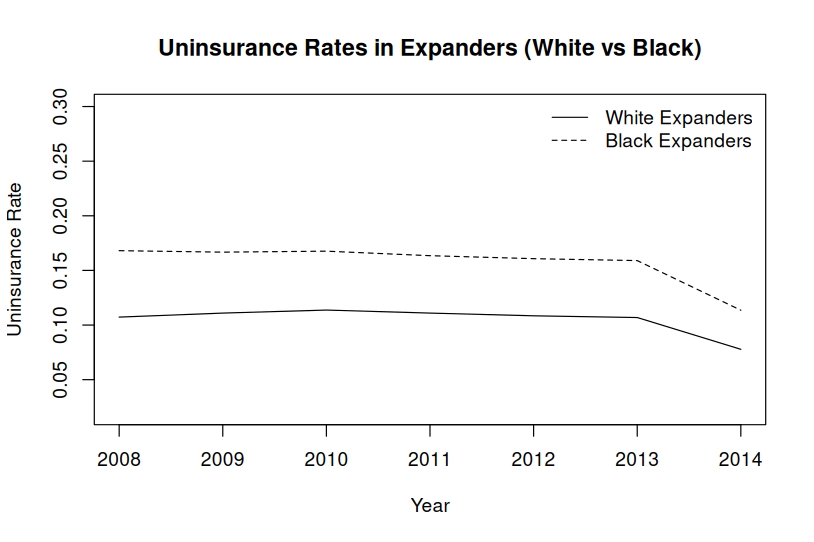}
\caption{}
  \end{subfigure}
  \caption{Uninsurance rates in expanding and non-expanding states, separately by race (panels (a) and (b)) and race-group-specific trends in expanding states (panel (c)). All would typically be taken to support the plausibility of relevant parallel trends assumptions.}
  \end{figure}\label{uninsurance_trends}
  
\section{Discussion}\label{section_discussion}
It is widely believed that a basic requirement for DiD to work is a data set comprising treated and control groups and pre- and post-treatment initiation periods. In this paper, we have presented some very simple yet possibly surprising facts about DiD formulas. As one example, in pre-post designs without a control group, the gDiD expression identifies effect modification of the ATT by group under a group `untreated' parallel trends assumption that is not stronger than that required by canonical DiD and also lends itself to pre-trends assessments. While the conditional ATTs are not identified in each group, effect heterogeneity can be of interest in its own right. In our running example application of \citet{kim2024synergistic}, effect modification by high UHC is suggestive that UHC might have been protective against harms from Covid. Another prime use case would be if equity is at issue. Consider the application of \citet{moriya2023racial} assessing the disparate effects of Medicaid expansion on health outcomes in Black and White individuals. If a control group were unavailable, one would still be able to target their estimand of interest using gDiD under a comparable assumption. Even if a control group were available, the group parallel trends assumption might still be more plausible and, again, could be partially assessed via pre-trends. We also provided graphical criteria based on SWIGs to potentially reject each of these assumptions. The criteria for pre-post `untreated' parallel trends depended in large part on whether exposure was shared across the population.

As a second example, even in the absence of an untreated pre-period in the treated group, the gDiD formula identifies variation in the ATT across time periods under the `untreated' parallel trends assumption. Thus, a program evaluator might be able to ascertain whether the impact of a program is diminishing over time using only data gathered from program participants and controls after the program was implemented, even if entry into the program was driven by unobserved variables that are prognostic for outcomes of interest and no instrumental variable is available. However, we emphasize that the group parallel trends assumption in this setting cannot be assessed via pre-trends and may lack structural justification due to contamination of counterfactual untreated trends in the study period by pre-baseline treatment.    

We further identified several data structures under which the `never switch' parallel trends assumption leads to identification of useful estimands. We discussed how this assumption can be motivated by a structural model (\ref{structural_never_switch}) similar to the traditional `untreated' assumption, and that structural model can often be partly assessed via pre-trends. We also provided a SWIG to illustrate that the `never switch' assumption requires similar substantive restrictions as canonical parallel trends, but somewhat stronger when observed treatment at the first time period differs between groups.

More generally, we have established that there is a large array of data structures and parallel trends assumptions under which the gDiD formula might identify useful estimands. One possible view is that this is purely good news. Canonical DiD is often considered a strong design, so now we can generate more strong evidence. Another possible view is that our results diminish canonical DiD by showing it to be an arbitrary member of a general class lacking its historical respectability. We take the view that any method is only as good as its underlying assumptions, which should ideally be evaluated on the merits as opposed to reputation. Researchers should carefully consider both the relevance of the estimand and plausibility of assumptions when selecting a study design. Tools to do this rigorously and reliably must continue to be developed. When pre-trends under relevant regimes are available, they seem a good place to start. We have also demonstrated how graphical reasoning with SWIGs can aid in the assessment of assumption plausibility.


In this paper, we focused almost entirely on identification. Future work should explore estimation and sensitivity analysis. While it would appear that results from canonical DiD should largely naturally port over to the alternative settings, subtleties may arise.

We are sure that gDiD estimates have been computed countless times by practitioners across the range of data structures we considered. They may often be arrived at either via solid common sense or in error. We hope we have helped to clarify their interpretation and justification. We also hope that these simple yet sometimes surprising results might inspire additional reassessments of well worn topics.

\section*{Acknowledgments}
We would like to acknowledge valuable research assistance from Alyssa Chen and Alex Ball. We also had helpful discussions with Clement de Chaisemartin.
\bibliography{bibliography}

\appendix
\section{Constructing graphical structural models} \label{section_swigs}
We build the maximal graph by beginning with the observable variables (in solid nodes).
We include split nodes for the treatment indicator $A_t$ at both time points $t \in \{0,1\}$.
The parallel trends assumption concerns outcomes under the regime ``no treatment in both periods'', 
so we intervene to set both equal to $0$. 
Then we also condition on $A_0 = 0$ (denoted by a rectangular node) because in this design, all units are untreated in period $t=0$.
Descendants of the split nodes are notated using corresponding potential outcomes, $Y_0(0)$ and $\Delta_1(0,0) = Y_1(0,0) - Y_0(0)$.

Then we consider relationships among the observed treatment and the solid nodes in the SWIG.
Because arrows from the right side of any split nodes will be conditioned on, we omit them from our assessment of d-separation.
But we do include arrows from baseline outcome to treatment at time $1$ (i.e., $Y_0 \rightarrow A_1$) and from baseline outcomes to the change in outcomes (i.e., $Y_0 \rightarrow \Delta_1$).
These are the only arrows we can add that respect the time ordering of the observable variables.

Next, we add nodes for unobservable common causes of the observed variables.
From the set $\{A_0,A_1,Y_0,\Delta_1\}$, we include unobserved common causes of each possible pair and triplet.
This is the maximum graph from which we then start removing arrows and nodes that will make the independence version of parallel trends true.

We begin with a pair of assumptions about relationships we are \emph{not} willing to remove;
these will constrain our subsequent choices.
First, we must be doing difference-in-differences for a reason, so we assume
\begin{equation}\label{eq:A2_cdid_swig}
\textbf{A1: } \mbox{There are unobserved common causes of treatment and baseline outcomes.}
\end{equation}
Thus, our graph must contain, $U_{A_1Y}$, $U_{A_0Y}$, and $U_{A_0A_1Y}$.
Second, we assume that outcomes are correlated,
\begin{equation}\label{eq:A3_cdid_swig}
\textbf{A2: } \mbox{There are unobserved common causes of baseline outcomes and untreated trends.}
\end{equation}
Thus, our graph must contain the node $U_{Y\Delta}$.

With these constraints, how can we obtain a graph with no backdoor paths from $A_1$ to $\Delta_1$?
First, we address the paths that go through unobservable common causes of $A$ and $\Delta_1$.
These provide backdoor paths that cannot be closed by conditioning because they go through unobservable nodes.
Several go only through unobservable nodes:
\begin{itemize}
\item $A_1 \leftarrow U_{A_1\Delta} \rightarrow \Delta$
\item $A_1 \leftarrow U_{A_0A_1\Delta} \rightarrow \Delta$
\item $A_1 \leftarrow U_{A_1Y\Delta} \rightarrow \Delta$
\end{itemize}
One goes through baseline treatment, which is a collider, but because it is conditioned on, this path is unblocked:
\begin{itemize}
\item $A_1 \leftarrow U_{A_0A_1} \rightarrow A_0 \leftarrow U_{A_0\Delta} \rightarrow \Delta$
\end{itemize}
To address these paths, we must remove $U_{A_0\Delta}$, $U_{A_1\Delta}$, $U_{A_0A_1\Delta}$, and $U_{A_1Y\Delta}$, which yields the following assumption:
\begin{equation}\label{eq:A1_cdid_swig}
\textbf{A3: } \mbox{There are no unobserved common causes of treatment and untreated trends.}
\end{equation}
This is perhaps how people might informally describe the parallel trends assumption.

We must also tackle the backdoor paths that go through $Y_0$. 
Some of these are paths on which $Y_0$ is a collider:
\begin{itemize}
\item $A_1 \leftarrow U_{A_1Y} \rightarrow Y_0 \leftarrow U_{Y\Delta} \rightarrow \Delta$
\item $A_1 \leftarrow U_{A_0A_1} \rightarrow A_0 \leftarrow U_{A_0Y} \rightarrow Y_0 \leftarrow U_{Y\Delta} \rightarrow \Delta$
\end{itemize}
while others go through $Y_0$ not as a collider:
\begin{itemize}
\item $A_1 \leftarrow Y_0 \rightarrow \Delta$
\item $A_1 \leftarrow Y_0 \leftarrow U_{Y\Delta} \rightarrow \Delta$
\item $A_1 \leftarrow U_{A_1Y} \rightarrow Y_0 \rightarrow \Delta$
\end{itemize}
Because are both collider and non-collider paths through $Y_0$, we cannot simply condition on $Y_0$.
Instead, we must get rid of the non-collider paths by removing arrows. 
Specifically, we must assume there is no $Y_0 \rightarrow A_1$ and no $Y_0 \rightarrow \Delta_1$.
This means that in addition to no \emph{unobserved} common causes of $A$ and $\Delta$, we also assume,
\begin{equation}\label{eq:A4_cdid_swig}
\textbf{A4: } \mbox{Baseline outcomes do not cause treatment or untreated trends.}
\end{equation}

\section{Derivations of Results in Table \ref{tab:array}} \label{section_proofs}
We provide derivations of the results summarized in Table 2 and describe each estimand in words.
\subsubsection*{Canonical, `Untreated' Parallel Trends}
\begin{align*}
    &E[Y_1-Y_0|G=1]-E[Y_1-Y_0|G=0]\\
    =&(E[Y_1-Y_0|G=1]-E[Y_1(0,0)-Y_0(0)|G=1])-\\
    &(E[Y_1-Y_0|G=0]-E[Y_1(0,0)-Y_0(0)|G=0])\\
    =&E[Y_1(0,1)-Y_1(0,0)|G=1]\\
    =&E[Y_1(0,1)-Y_1(0,0)|A_1=1].
\end{align*}
This is the ATT at the second time point.

\subsubsection*{Pre-post, `Untreated' Parallel Trends}
\begin{align*}
    &E[Y_1-Y_0|G=1] - E[Y_1-Y_0|G=0]\\
=&(E[Y_1-Y_0|G=1]-E[Y_1(0,0)-Y_0(0)|G=1]) - \\
&(E[Y_1-Y_0|G=0]-E[Y_1(0,0)-Y_0(0)|G=0])\\
=&(E[Y_1(0,1)-Y_0(0)|G=1]-E[Y_1(0,0)-Y_0(0)|G=1]) - \\
&(E[Y_1(0,1)-Y_0(0)|G=0]-E[Y_1(0,0)-Y_0(0)|G=0])\\
=&E[Y_1(0,1)-Y_1(0,0)|G=1] - E[Y_1(0,1)-Y_1(0,0)|G=0]
\end{align*}
This is the difference between the effects in $G=1$ and $G=0$ of starting treatment at the second time point.

\subsubsection*{No Pre-period, Untreated Parallel Trends}
We consider treatment as a fixed point exposure in this setting.
\begin{align*}
    &E[Y_1-Y_0|G=1] - E[Y_1-Y_0|G=0]\\
=&(E[Y_1-Y_0|G=1]-E[Y_1(0)-Y_0(0)|G=1]) - \\
&(E[Y_1-Y_0|G=0]-E[Y_1(0)-Y_0(0)|G=0])\\
=&(E[Y_1(1)-Y_0(1)|G=1]-E[Y_1(0)-Y_0(0)|G=1]) - \\
&(E[Y_1(0)-Y_0(0)|G=0]-E[Y_1(0)-Y_0(0)|G=0])\\
=&E[Y_1(1)-Y_1(0)|G=1] - E[Y_0(1)-Y_0(0)|G=1]\\
=&E[Y_1(1)-Y_1(0)|A=1] - E[Y_0(1)-Y_0(0)|A=1]
\end{align*}
This is the difference between the effects at the second time point and the first time point in the $G=A=1$ group.

\subsubsection*{Treatment Turns On in $G=1$ vs. Treated, Untreated Parallel Trends}
\begin{align*}
    &E[Y_1-Y_0|G=1] - E[Y_1-Y_0|G=0]\\
=&(E[Y_1-Y_0|G=1]-E[Y_1(0,0)-Y_0(0)|G=1]) - \\
&(E[Y_1-Y_0|G=0]-E[Y_1(0,0)-Y_0(0)|G=0])\\
=&E[Y_1(0,1)-Y_1(0,0)|G=1] - \\
&(E[Y_1(1,1)-Y_1(0,0)|G=0]-E[Y_0(1)-Y_0(0)|G=0])
\end{align*}
This is the difference between the effect of treatment at the second time point in the $G=1$ group and the difference between the total effect of treatment at both times on the outcome at the second time and the effect of treatment at the first time in the $G=0$ group.

\subsubsection*{Treatment Turns Off in $G=1$ vs. Treated, Untreated Parallel Trends}
\begin{align*}
    &E[Y_1-Y_0|G=1] - E[Y_1-Y_0|G=0]\\
=&(E[Y_1-Y_0|G=1]-E[Y_1(0,0)-Y_0(0)|G=1]) - \\
&(E[Y_1-Y_0|G=0]-E[Y_1(0,0)-Y_0(0)|G=0])\\
=&(E[Y_1(1,0)-Y_1(0,0)|G=1]-E[Y_0(1)-Y_0(0)]) - \\
&(E[Y_1(1,1)-Y_0(0,0)|G=0]-E[Y_0(1)-Y_0(0)|G=0])
\end{align*}
This is the difference between the lasting effect of treatment at the first time point on the outcome at the second time point and the immediate effect of treatment at the first time point in the $G=1$ group subtracted by the difference between the total effect of treatment at both times on the outcome at the second time and the immediate effect of treatment at the first time in the $G=0$ group. The estimand does not simplify meaningfully under a No Carryover assumption.

\subsubsection*{Treatment Turns Off in $G=1$ vs. Untreated, Untreated Parallel Trends}
\begin{align*}
    &E[Y_1-Y_0|G=1] - E[Y_1-Y_0|G=0]\\
=&(E[Y_1-Y_0|G=1]-E[Y_1(0,0)-Y_0(0)|G=1]) - \\
&(E[Y_1-Y_0|G=0]-E[Y_1(0,0)-Y_0(0)|G=0])\\
=&(E[Y_1(1,0)-Y_1(0,0)|G=1]-E[Y_0(1)-Y_0(0)])
\end{align*}
This is the difference between the lasting effect of treatment at the first time point on the outcome at the second time point and the immediate effect of treatment at the first time point in the $G=1$ group. Under a No Carryover assumption, it is (negative) the immediate effect of treatment at the first time point in the $G=1$ group.

\subsubsection*{Crossover, Untreated Parallel Trends}
\begin{align*}
    &E[Y_1-Y_0|G=1] - E[Y_1-Y_0|G=0]\\
=&(E[Y_1-Y_0|G=1]-E[Y_1(0,0)-Y_1(0)|G=1]) - \\&
(E[Y_1-Y_0|G=0]-E[Y_1(0,0)-Y_0(0)|G=0])\\
=&(E[Y_1(1,0)-Y_1(0,0)|G=1] - E[Y_0(1)-Y_0(0)|G=1])-\\
& E[Y_1(0,1)-Y_0(0,0)|G=0]
\end{align*}
This is the difference between the delayed effect of treatment at the first time point on the outcome at the second time point and the immediate effect of treatment at the first time point in the $G=1$ group subtracted by the immediate effect of treatment at the second time point in the $G=0$ group. Under a No Carryover assumption, the effect simplifies to
\begin{equation*}
    E[Y_1(0,1)-Y_0(0,0)|G=0]- E[Y_0(1)-Y_0(0)|G=1],
\end{equation*}
which is the difference between the effect of treatment at the second time point in the $G=0$ group and the effect of treatment at the first time point in the $G=1$ group.

\subsubsection*{Switch off Pre-post, Untreated Parallel Trends}
\begin{align*}
    &E[Y_1-Y_0|G=1] - E[Y_1-Y_0|G=0]\\
=&(E[Y_1-Y_0|G=1]-E[Y_1(0,0)-Y_1(0)|G=1]) - \\
&(E[Y_1-Y_0|G=0]-E[Y_1(0,0)-Y_0(0)|G=0])\\
=& (E[Y_1(1,0)-Y_1(0,0)|G=1]-E[Y_0(1)-Y_0(0)|G=1]) - \\
&(E[Y_1(1,0)-Y_1(0,0)|G=0]-E[Y_0(1)-Y_0(0)|G=0])
\end{align*}
This is the difference between the long term and immediate effects of treatment at the first time point in the $G=1$ group subtracted by the difference between the long term and immediate effects of treatment at the first time point in the $G=0$ group. Does the effect decay more in one group than another? Under a No Carryover assumption, this reduces to 
\begin{equation*}
    E[Y_0(1)-Y_0(0)|G=0]-E[Y_0(1)-Y_0(0)|G=1],
\end{equation*}
which is the difference between the immediate effect of treatment at the first time point in the two groups.

\subsubsection*{Canonical, Treated Parallel Trends}
\begin{align*}
    &E[Y_1-Y_0|G=1]-E[Y_1-Y_0|G=0]\\
    =&(E[Y_1-Y_0|G=1]-E[Y_1(0,0)-Y_0(0)|G=1])-\\
    &(E[Y_1-Y_0|G=0]-E[Y_1(0,0)-Y_0(0)|G=0])\\
    =&(E[Y_1(0,1)-Y_0(0)|G=1]-E[Y_1(1,1)-Y_0(1)|G=1])-\\
    &(E[Y_1(0,0)-Y_0(0)|G=0]-E[Y_1(1,1)-Y_0(1)|G=0])\\
    =&(E[Y_1(0,1)-Y_1(1,1)|G=1]-E[Y_0(0)-Y_0(1)|G=1])-\\
    &(E[Y_1(0,0)-Y_1(1,1)|G=0]-E[Y_0(0)-Y_0(1)|G=0])
\end{align*}
This is the difference between effect of no treatment followed by treatment compared to treatment at both times on the outcome at the second time point and the immediate effect of no treatment compared to treatment at the first time point in the $G=1$ group subtracted by the difference between the effect of never treatment compared to always treatment on the outcome at the second time point and the immediate effect of no treatment compared to treatment at the first time point in the $G=0$ group. The estimand does not simplify meaningfully under a No Carryover assumption.

\subsubsection*{Pre-post, Treated Parallel Trends}
\begin{align*}
    &E[Y_1-Y_0|G=1] - E[Y_1-Y_0|G=0]\\
=&(E[Y_1-Y_0|G=1]-E[Y_1(0)-Y_0(0)|G=1]) - \\
&(E[Y_1-Y_0|G=0]-E[Y_1(0)-Y_0(0)|G=0])\\
=&(E[Y_1(0,1)-Y_1(1,1)|G=1]-E[Y_0(0)-Y_0(1)|G=1]) - \\
&(E[Y_1(0,1)-Y_1(1,1)|G=0]-E[Y_0(0)-Y_0(1)|G=0])
\end{align*}
This is the difference between the effect of no treatment followed by treatment compared to treatment at both times on the outcome at the second time point and the immediate effect of no treatment compared to treatment at the first time point in the $G=1$ group subtracted by the same difference in the $G=0$ group. Under a No Carryover assumption, this simplifies to
\begin{equation*}
    E[Y_0(1)-Y_0(0)|G=1]-E[Y_0(1)-Y_0(0)|G=0],
\end{equation*}
which is the difference between immediate effects of treatment at the first time period in the two groups.

\subsubsection*{No Pre-period ($G=A=1$), Treated Parallel Trends}
We consider treatment as a fixed point exposure in this setting.
\begin{align*}
    &E[Y_1-Y_0|G=1] - E[Y_1-Y_0|G=0]\\
=&(E[Y_1-Y_0|G=1]-E[Y_1(1)-Y_0(1)|G=1]) - \\
&(E[Y_1-Y_0|G=0]-E[Y_1(1)-Y_0(1)|G=0])\\
=&(E[Y_1(1)-Y_0(1)|G=1]-E[Y_1(1)-Y_0(1)|G=1]) - \\
&(E[Y_1(0)-Y_0(0)|G=0]-E[Y_1(1)-Y_0(1)|G=0])\\
=&E[Y_1(1)-Y_1(0)|G=0] - E[Y_0(1)-Y_0(0)|G=0]\\
=&E[Y_1(1)-Y_1(0)|A=0] - E[Y_0(1)-Y_0(0)|A=0]
\end{align*}
This is the difference between the effects at the second time point and the first time point in the $G=A=0$ group.

\subsubsection*{Treatment Turns On in $G=1$ vs. Treated, Treated Parallel Trends}
\begin{align*}
    &E[Y_1-Y_0|G=1] - E[Y_1-Y_0|G=0]\\
=&(E[Y_1-Y_0|G=1]-E[Y_1(1,1)-Y_0(1)|G=1]) - \\
&(E[Y_1-Y_0|G=0]-E[Y_1(1,1)-Y_0(1)|G=0])\\
=&E[Y_1(0,1)-Y_1(1,1)|G=1]-E[Y_0(0)-Y_0(1)|G=1]
\end{align*}
This is the difference between the lasting (one time step ahead) and immediate effects of withholding treatment in the $G=1$ group. Under a No Carryover assumption, the estimand simplifies to 
\begin{equation*}
    E[Y_0(1)-Y_0(0)|G=1],
\end{equation*}
which is the immediate effect of treatment in the first time point in the $G=1$ group.

\subsubsection*{Treatment Turns Off in $G=1$ vs. Treated, Treated Parallel Trends}
\begin{align*}
    &E[Y_1-Y_0|G=1] - E[Y_1-Y_0|G=0]\\
=&(E[Y_1-Y_0|G=1]-E[Y_1(1,1)-Y_0(1)|G=1]) - \\
&(E[Y_1-Y_0|G=0]-E[Y_1(1,1)-Y_0(1)|G=0])\\
=&E[Y_1(1,0)-Y_1(1,1)|G=1]
\end{align*}
This is the effect of withholding treatment at the second time point in the $G=1$ group.

\subsubsection*{Treatment Turns Off in $G=1$ vs. Untreated, Treated Parallel Trends}
\begin{align*}
    &E[Y_1-Y_0|G=1] - E[Y_1-Y_0|G=0]\\
=&(E[Y_1-Y_0|G=1]-E[Y_1(1,1)-Y_0(1)|G=1]) - \\
&(E[Y_1-Y_0|G=0]-E[Y_1(1,1)-Y_0(1)|G=0])\\
=&E[Y_1(1,0)-Y_1(1,1)|G=1]-\\
&(E[Y_1(0,0)- Y_1(1,1)|G=0]-E[Y_0(0)-Y_0(1)|G=0])
\end{align*}
This is the difference between the effect of withholding treatment at the second time point in the $G=1$ group and the difference between the . The estimand does not simplify meaningfully under a No Carryover assumption.

\subsubsection*{Crossover, Treated Parallel Trends}
\begin{align*}
    &E[Y_1-Y_0|G=1] - E[Y_1-Y_0|G=0]\\
=&(E[Y_1-Y_0|G=1]-E[Y_1(1,1)-Y_1(1)|G=1]) - \\
&(E[Y_1-Y_0|G=0]-E[Y_1(1,1)-Y_0(1)|G=0])\\
=&E[Y_1(1,0)-Y_1(1,1)|G=1] -\\
&(E[Y_1(0,1)-Y_1(1,1)|G=0]-E[Y_0(0)-Y_0(1)|G=0])
\end{align*}
This is the immediate effect of withholding treatment at the second time point in the $G=1$ group subtracted by the difference between the delayed effect of withholding treatment at the first time point on the outcome at the second time point and the immediate effect of withholding treatment at the first time point in the $G=0$ group. Under a No Carryover assumption, the effect simplifies to
\begin{equation*}
    E[Y_0(0)-Y_0(1)|G=0]-E[Y_1(0)-Y_1(1)|G=1], 
\end{equation*}
which is the difference between the effect of treatment at the first time point in the $G=0$ group and the effect of treatment at the second time point in the $G=1$ group.

\subsubsection*{Canonical, Untreated then Treated Parallel Trends}
\begin{align*}
    &E[Y_1-Y_0|G=1]-E[Y_1-Y_0|G=0]\\
    =&(E[Y_1-Y_0|G=1]-E[Y_1(0,1)-Y_0(0)|G=1])-\\
    &(E[Y_1-Y_0|G=0]-E[Y_1(0,1)-Y_0(0)|G=0])\\
    =&E[Y_1(0,0)-Y_1(0,1)|G=0]\\
\end{align*}
This is the immediate effect of withholding treatment at the second time point in the $G=0$ group. 

\subsubsection*{Pre-post, Untreated then Treated Parallel Trends}
\begin{align*}
    &E[Y_1-Y_0|G=1] - E[Y_1-Y_0|G=0]\\
=&(E[Y_1-Y_0|G=1]-E[Y_1(0,1)-Y_0(0)|G=1]) - \\
&(E[Y_1-Y_0|G=0]-E[Y_1(0,1)-Y_0(0)|G=0])\\
=&0
\end{align*}
Everything cancels.

\subsubsection*{No Pre-period ($G=A=1$), Untreated then Treated Parallel Trends}
\begin{align*}
    &E[Y_1-Y_0|G=1] - E[Y_1-Y_0|G=0]\\
=&(E[Y_1-Y_0|G=1]-E[Y_1(0,1)-Y_0(0)|G=1]) - \\
&(E[Y_1-Y_0|G=0]-E[Y_1(0,1)-Y_0(0)|G=0])\\
=&(E[Y_1(1,1)-Y_1(0,1)|G=1]-E[Y_0(1)-Y_0(0)|G=1]) -\\
&E[Y_1(0,0)-Y_1(0,1)|G=0]
\end{align*}
This is the difference between the effect of early treatment initiation on the outcome at the second time point and the immediate effect of initiating treatment at the first time point in the $G=1$ group subtracted by the effect of withholding treatment at the second time point in the $G=0$ group. Under a No Carryover assumption, this simplifies to 
\begin{equation*}
   E[Y_1(0,1)-Y_1(0,0)|G=0] -E[Y_0(1)-Y_0(0)|G=1],
\end{equation*}
which is the immediate effect of treatment at the second time point in the $G=0$ group subtracted by the immediate effect of treatment at the first time point in the $G=1$ group.

\subsubsection*{Treatment Turns On in $G=1$ vs. Treated, Switch On Parallel Trends}
\begin{align*}
    &E[Y_1-Y_0|G=1] - E[Y_1-Y_0|G=0]\\
=&(E[Y_1-Y_0|G=1]-E[Y_1(0,1)-Y_0(0)|G=1]) - \\
&(E[Y_1-Y_0|G=0]-E[Y_1(0,1)-Y_0(0)|G=0])\\
=&E[Y_1(1,1)-Y_1(0,1)|G=0] - E[Y_0(1)-Y_0(0)|G=0]
\end{align*}
This is the effect of early initiation compared to late initiation on the outcome at the second time point subtracted by the immediate effect of early initiation in the $G=0$ group

\subsubsection*{Treatment Turns Off in $G=1$ vs. Treated, Untreated then Treated Parallel Trends}
\begin{align*}
    &E[Y_1-Y_0|G=1] - E[Y_1-Y_0|G=0]\\
=&(E[Y_1-Y_0|G=1]-E[Y_1(0,1)-Y_0(0)|G=1]) - \\
&(E[Y_1-Y_0|G=0]-E[Y_1(0,1)-Y_0(0)|G=0])\\
=&(E[Y_1(1,0)-Y_1(0,1)|G=1]-E[Y_0(1)-Y_0(0)|G=1])-\\
&E[Y_1(0,0)-Y_1(0,1)|G=0]
\end{align*}
This is the difference between the effect of starting then stopping treatment compared to starting treatment at the second time point on the outcome at the second time point in the $G=1$ group subtracted by the effect of continuing to withhold treatment at the second time point on the outcome at the second time point in the $G=0$ group. It does not simplify meaningfully under a No Carryover assumption.

\subsubsection*{Treatment Turns Off in $G=1$ vs. Untreated, Untreated then Treated Parallel Trends}
\begin{align*}
    &E[Y_1-Y_0|G=1] - E[Y_1-Y_0|G=0]\\
=&(E[Y_1-Y_0|G=1]-E[Y_1(0,1)-Y_0(0)|G=1]) - \\
&(E[Y_1-Y_0|G=0]-E[Y_1(0,1)-Y_0(0)|G=0])\\
=&(E[Y_1(1,0)-Y_1(0,1)|G=1]-E[Y_1(1)-Y_0(0)|G=1])-\\
&E[Y_1(0,0)-Y_1(0,1)|G=0]
\end{align*}
This is the difference between the effect of starting then stopping compared to withholding then starting treatment on the outcome at the second time point and the immediate effect of starting treatment at the first time point in the $G=1$ group subtracted by the effect of continuing to withhold treatment at the second time point in the $G=0$ group. The estimand does not simplify meaningfully under a No Carryover assumption.

\subsubsection*{Crossover, Untreated then Treated Parallel Trends}
\begin{align*}
    &E[Y_1-Y_0|G=1] - E[Y_1-Y_0|G=0]\\
=&(E[Y_1-Y_0|G=1]-E[Y_1(0,1)-Y_1(0)|G=1]) - \\
&(E[Y_1-Y_0|G=0]-E[Y_1(0,1)-Y_0(0)|G=0])\\
=&E[Y_1(1,0)-Y_1(0,1)|G=1]-E[Y_0(1)-Y_0(0)|G=1]
\end{align*}
This is the difference between the effect of starting then stopping compared to withholding then starting treatment on the outcome at the second time point and the immediate effect of starting treatment at the first time point in the $G=1$ group. Under a No Carryover assumption, the effect simplifies to
\begin{equation*}
    E[Y_1(0)-E[Y_1(1)|G=1]-E[Y_0(1)-Y_0(0)|G=1], 
\end{equation*}
which is the difference between the effect of withholding treatment at the second time point and giving treatment at the first time point in the $G=1$ group.

\subsubsection*{Switch off Pre-post, Untreated then Treated Parallel Trends}
\begin{align*}
    &E[Y_1-Y_0|G=1] - E[Y_1-Y_0|G=0]\\
=&(E[Y_1-Y_0|G=1]-E[Y_1(0,1)-Y_1(0)|G=1]) - \\
&(E[Y_1-Y_0|G=0]-E[Y_1(0,1)-Y_0(0)|G=0])\\
=&(E[Y_1(1,0)-Y_1(0,1)|G=1]-E[Y_0(1)-Y_0(0)|G=1]) - \\
&(E[Y_1(1,0)-Y_1(0,1)|G=0]-E[Y_0(1)-Y_0(0)|G=0])
\end{align*}
Under a No Carryover assumption, this reduces to 
\begin{equation*}
    (E[Y_1(0)-Y_1(1)|G=1]+E[Y_0(0)-Y_0(1)|G=1]) - (E[Y_1(0)-Y_1(1)|G=0]+E[Y_0(0)-Y_0(1)|G=0]),
\end{equation*}
which is the difference between the sums of the immediate effects of withholding treatment in each group.

\subsubsection*{Canonical, Treated then Untreated Parallel Trends}
\begin{align*}
    &E[Y_1-Y_0|G=1]-E[Y_1-Y_0|G=0]\\
    =&(E[Y_1-Y_0|G=1]-E[Y_1(1,0)-Y_0(1)|G=1])-\\
    &(E[Y_1-Y_0|G=0]-E[Y_1(1,0)-Y_0(1)|G=0])\\
    =&(E[Y_1(0,1)-Y_1(1,0)|G=1]-E[Y_0(0)-Y_0(1)|G=1])-\\
    &(E[Y_1(0,0)-Y_1(1,0)|G=0]-E[Y_0(0)-Y_0(1)|G=0])\\
\end{align*}
This the difference between the effect of initiating treatment at the second time point compared to stopping treatment at the second time point and the immediate effect of withholding treatment at the first time point in the $G=1$ group subtracted by the difference between the effect of withholding treatment at both times compared to treating only at the first time and the immediate effect of withholding treatment at the first time point in the $G=0$ group. It does not simplify meaningfully under a No Carryover assumption.

\subsubsection*{Pre-post, Treated then Untreated Parallel Trends}
\begin{align*}
    &E[Y_1-Y_0|G=1] - E[Y_1-Y_0|G=0]\\
&=(E[Y_1-Y_0|G=1]-E[Y_1(1,0)-Y_0(1)|G=1]) - \\
&(E[Y_1-Y_0|G=0]-E[Y_1(1,0)-Y_0(1)|G=0])\\
&=(E[Y_1(0,1)-Y_1(1,0)|G=1]-E[Y_0(0)-Y_0(1)|G=1]) - \\
&(E[Y_1(0,1)-Y_1(1,0)|G=0]-E[Y_0(0)-Y_0(1)|G=0])
\end{align*}
This the difference between the effect of initiating treatment at the second time point compared to stopping treatment at the second time point and the immediate effect of withholding treatment at the first time point in the $G=1$ group subtracted by the same difference in the $G=0$ group.
Under No Carryover, this reduces to 
\begin{equation}
    E[Y_1(1)-Y_1(0)|G=1]+E[Y_0(1)-Y_0(0)|G=1]) - (E[Y_1(1)-Y_1(0)|G=0]+E[Y_0(1)-Y_0(0)|G=0]),
\end{equation}
which is the difference between the sums of the immediate effects at each time point in the two groups.

\subsubsection*{No Pre-period ($G=A=1$), Treated then Untreated Parallel Trends}
\begin{align*}
    &E[Y_1-Y_0|G=1] - E[Y_1-Y_0|G=0]\\
&=(E[Y_1-Y_0|G=1]-E[Y_1(1,0)-Y_0(1)|G=1]) - \\
&(E[Y_1-Y_0|G=0]-E[Y_1(1,0)-Y_0(1)|G=0])\\
&=E[Y_1(1,1)-Y_1(1,0)|G=1] - (E[Y_1(0,0)-Y_1(1,0)|G=0]-E[Y_0(0)-Y_0(1)|G=0])
\end{align*}
Under a No Carryover assumption, this simplifies to 
\begin{equation*}
   E[Y_1(1)-Y_1(0)|G=1] - E[Y_0(1)-Y_0(0)|G=0],
\end{equation*}
which is the immediate effect of treatment at the second time point in the $G=1$ group subtracted by the immediate effect of treatment at the first time point in the $G=1$ group.

\subsubsection*{Treatment Turns On in $G=1$ vs. Treated, Treated then Untreated Parallel Trends}
\begin{align*}
    &E[Y_1-Y_0|G=1] - E[Y_1-Y_0|G=0]\\
&=(E[Y_1-Y_0|G=1]-E[Y_1(1,0)-Y_0(1)|G=1]) - \\
&(E[Y_1-Y_0|G=0]-E[Y_1(1,0)-Y_0(1)|G=0])\\
&=(E[Y_1(0,1)-Y_1(1,0)|G=1]-E[Y_0(0)-Y_0(1)|G=1]) - E[Y_1(1,1)-Y_1(1,0)|G=0]
\end{align*}
This is the difference between the effect of starting treatment at the second time point compared to stopping treatment at the second time point on the outcome at the second time point in the $G=1$ group subtracted by the effect of continuing versus stopping treatment at the second time point in the $G=0$ group. It does not meaningfully simplify under a No Carryover assumption.

\subsubsection*{Treatment Turns Off in $G=1$ vs. Treated, Treated then Untreated Parallel Trends}
\begin{align*}
    &E[Y_1-Y_0|G=1] - E[Y_1-Y_0|G=0]\\
&=(E[Y_1-Y_0|G=1]-E[Y_1(1,0)-Y_0(1)|G=1]) - \\
&(E[Y_1-Y_0|G=0]-E[Y_1(1,0)-Y_0(1)|G=0])\\
&= E[Y_1(1,0)-Y_1(1,1)|G=0]
\end{align*}
This is the effect of switching treatment off at the second time point in the $G=0$ group.

\subsubsection*{Treatment Turns Off in $G=1$ vs. Untreated, Treated then Untreated Parallel Trends}
\begin{align*}
    &E[Y_1-Y_0|G=1] - E[Y_1-Y_0|G=0]\\
&=(E[Y_1-Y_0|G=1]-E[Y_1(1,0)-Y_0(1)|G=1]) - \\
&(E[Y_1-Y_0|G=0]-E[Y_1(1,0)-Y_0(1)|G=0])\\
&= E[Y_0(0)-Y_0(1)|G=0]-E[Y_1(0,0)-Y_0(1,0)|G=0]
\end{align*}
This is the difference between the immediate effect of withholding treatment at the first time point and the long term effect of withholding treatment at the first time point in the $G=0$ group. Under a No Carryover assumption, it simplifies to
\begin{equation*}
    E[Y_0(0)-Y_0(1)|G=0],
\end{equation*}
which is the immediate effect of withholding treatment at the first time point in the $G=0$ group.

\subsubsection*{Crossover, Treated then Untreated Parallel Trends}
\begin{align*}
    &E[Y_1-Y_0|G=1] - E[Y_1-Y_0|G=0]\\
&=(E[Y_1-Y_0|G=1]-E[Y_1(1,0)-Y_1(1)|G=1]) - \\
&(E[Y_1-Y_0|G=0]-E[Y_1(1,0)-Y_0(1)|G=0])\\
&= E[Y_0(0)-Y_0(1)|G=0]-E[Y_1(0,1)-Y_0(1,0)|G=0]
\end{align*}
Under a No Carryover assumption, the effect simplifies to
\begin{equation*}
    E[Y_0(0)-Y_0(1)|G=0]-E[Y_1(1)-Y_0(0)|G=0], 
\end{equation*}
which is the difference between the effect of withholding treatment at the first time point and the effect of giving treatment at the second time point in the $G=0$ group.

\subsubsection*{Canonical, Never Switch Parallel Trends}
\begin{align*}
    &E[Y_1-Y_0|G=1]-E[Y_1-Y_0|G=0]\\
    &=(E[Y_1-Y_0|G=1]-E[Y_1(0,0)-Y_0(0)|G=1])-\\
    &(E[Y_1-Y_0|G=0]-E[Y_1(0,0)-Y_0(0)|G=0])\\
    &=(E[Y_1(0,1)-Y_1(0,0)|G=1]
\end{align*}
This is the same ATT in $G=1$ as the standard untreated parallel trends assumption.

\subsubsection*{Pre-post, Never Switch Parallel Trends}
\begin{align*}
    &E[Y_1-Y_0|G=1] - E[Y_1-Y_0|G=0]\\
&=(E[Y_1-Y_0|G=1]-E[Y_1(0,0)-Y_0(0)|G=1]) - \\
&(E[Y_1-Y_0|G=0]-E[Y_1(0,0)-Y_0(0)|G=0])\\
&=E[Y_1(0,1)-Y_1(0,0)|G=1]-E[Y_1(0,1)-Y_1(0,0)|G=0]
\end{align*}
This is the difference between the effect of initiating treatment at the second time point in the two groups.

\subsubsection*{No Pre-period ($G=A=1$), Never Switch Parallel Trends}
\begin{align*}
    &E[Y_1-Y_0|G=1] - E[Y_1-Y_0|G=0]\\
&=(E[Y_1-Y_0|G=1]-E[Y_1(1,1)-Y_0(1)|G=1]) - \\
&(E[Y_1-Y_0|G=0]-E[Y_1(0,0)-Y_0(0)|G=0])\\
&=0
\end{align*}

\subsubsection*{Treatment Turns On in $G=1$ vs. Treated, Never Switch Parallel Trends}
\begin{align*}
    &E[Y_1-Y_0|G=1] - E[Y_1-Y_0|G=0]\\
&=(E[Y_1-Y_0|G=1]-E[Y_1(0,0)-Y_0(0)|G=1]) - \\
&(E[Y_1-Y_0|G=0]-E[Y_1(1,1)-Y_0(1)|G=0])\\
&=E[Y_1(0,1)-Y_1(0,0)|G=1] 
\end{align*}
This is the effect of starting treatment at the second time point in the $G=1$ group.

\subsubsection*{Treatment Turns Off in $G=1$ vs. Treated, Never Switch Parallel Trends}
\begin{align*}
    &E[Y_1-Y_0|G=1] - E[Y_1-Y_0|G=0]\\
&=(E[Y_1-Y_0|G=1]-E[Y_1(1,1)-Y_0(1)|G=1]) - \\
&(E[Y_1-Y_0|G=0]-E[Y_1(1,1)-Y_0(1)|G=0])\\
&= E[Y_1(1,0)-Y_1(1,1)|G=1]
\end{align*}
This is the effect of switching treatment off at the second time point in the $G=1$ group.

\subsubsection*{Treatment Turns Off in $G=1$ vs. Untreated, Never Switch Parallel Trends}
\begin{align*}
    &E[Y_1-Y_0|G=1] - E[Y_1-Y_0|G=0]\\
&=(E[Y_1-Y_0|G=1]-E[Y_1(1,1)-Y_0(1)|G=1]) - \\
&(E[Y_1-Y_0|G=0]-E[Y_1(0,0)-Y_0(0)|G=0])\\
&= E[Y_1(1,0)-Y_0(1,1)|G=1]
\end{align*}
This is the effect of switching treatment off at the second time point in the $G=1$ group. 

\subsubsection*{Crossover, Never Switch Parallel Trends}
\begin{align*}
    &E[Y_1-Y_0|G=1] - E[Y_1-Y_0|G=0]\\
&=(E[Y_1-Y_0|G=1]-E[Y_1(1,1)-Y_1(1)|G=1]) - \\
&(E[Y_1-Y_0|G=0]-E[Y_1(0,0)-Y_0(0)|G=0])\\
&= E[Y_1(1,0)-Y_1(1,1)|G=1] - E[Y_1(0,1)-Y_1(0,0)|G=0]
\end{align*}
This is the difference between the effects of switching treatment at the second time point in the two groups.

\subsubsection*{Canonical, Always Switch Parallel Trends}
\begin{align*}
    &E[Y_1-Y_0|G=1]-E[Y_1-Y_0|G=0]\\
    &=(E[Y_1-Y_0|G=1]-E[Y_1(0,1)-Y_0(0)|G=1])-\\
    &(E[Y_1-Y_0|G=0]-E[Y_1(0,1)-Y_0(0)|G=0])\\
    &=E[Y_1 (0,0)-Y_1 (0,1)|G=0]
\end{align*}
This is the effect of withholding treatment at the second time point in the $G=0$ group.

\subsubsection*{Pre-post, Always Switch Parallel Trends}
\begin{align*}
    &E[Y_1-Y_0|G=1] - E[Y_1-Y_0|G=0]\\
&=(E[Y_1-Y_0|G=1]-E[Y_1(0,1)-Y_0(0)|G=1]) - \\
&(E[Y_1-Y_0|G=0]-E[Y_1(0,1)-Y_0(0)|G=0])\\
&=0
\end{align*}
Everything cancels

\subsubsection*{No Pre-period ($G=A=1$), Always Switch Parallel Trends}
\begin{align*}
    &E[Y_1-Y_0|G=1] - E[Y_1-Y_0|G=0]\\
&=(E[Y_1-Y_0|G=1]-E[Y_1(1,0)-Y_0(1)|G=1]) - \\
&(E[Y_1-Y_0|G=0]-E[Y_1(0,1)-Y_0(0)|G=0])\\
&=E[Y_1(1,1)-Y_1(1,0)|G=1]-E[Y_0(0,0)-Y_1(0,1)|G=0]
\end{align*}
This is the effect of continuing treatment at the second time point in $G=1$ subtracted by the effect of continuing to withhold treatment at the second time point in the $G=0$ group. Which group gained more by staying the course?

\subsubsection*{Treatment Turns On in $G=1$ vs. Treated, Always Switch Parallel Trends}
\begin{align*}
    &E[Y_1-Y_0|G=1] - E[Y_1-Y_0|G=0]\\
&=(E[Y_1-Y_0|G=1]-E[Y_1(0,1)-Y_0(0)|G=1]) - \\
&(E[Y_1-Y_0|G=0]-E[Y_1(1,0)-Y_0(1)|G=0])\\
&= E[Y_1(1,0)-Y_1(1,1)|G=0] 
\end{align*}
This is the effect of switching treatment off at the second time point in the $G=0$ group.

\subsubsection*{Treatment Turns Off in $G=1$ vs. Treated, Always Switch Parallel Trends}
\begin{align*}
    &E[Y_1-Y_0|G=1] - E[Y_1-Y_0|G=0]\\
&=(E[Y_1-Y_0|G=1]-E[Y_1(1,0)-Y_0(1)|G=1]) - \\
&(E[Y_1-Y_0|G=0]-E[Y_1(1,0)-Y_0(1)|G=0])\\
&= E[Y_1(1,0)-Y_1(1,1)|G=0]
\end{align*}
This is the effect of switching treatment off at the second time point in the $G=0$ group.

\subsubsection*{Treatment Turns Off in $G=1$ vs. Untreated, Always Switch Parallel Trends}
\begin{align*}
    &E[Y_1-Y_0|G=1] - E[Y_1-Y_0|G=0]\\
&=(E[Y_1-Y_0|G=1]-E[Y_1(1,0)-Y_0(1)|G=1]) - \\
&(E[Y_1-Y_0|G=0]-E[Y_1(0,1)-Y_0(0)|G=0])\\
&= E[Y_0(0,1)-Y_1(0,0)|G=0]
\end{align*}
This is the effect of starting treatment at the second time point in the $G=0$ group. 

\subsubsection*{Crossover, Always Switch Parallel Trends}
\begin{align*}
    &E[Y_1-Y_0|G=1] - E[Y_1-Y_0|G=0]\\
&=(E[Y_1-Y_0|G=1]-E[Y_1(1,0)-Y_1(1)|G=1]) - \\
&(E[Y_1-Y_0|G=0]-E[Y_1(0,1)-Y_0(0)|G=0])\\
&= 0
\end{align*}
Everything cancels 

\end{document}